\pdfoutput=1
\documentclass[a4paper,11pt]{article}
\usepackage{amsmath,amssymb,amsfonts,amscd}
\usepackage[multiple,stable]{footmisc}
\usepackage[amsmath,amsthm,thmmarks]{ntheorem}
\usepackage{rotating}
\allowdisplaybreaks[4]
\usepackage[noconfig]{refstyle}
\usepackage[dvipsnames]{xcolor}
\usepackage[hyperfootnotes=false, linktocpage=true, colorlinks, citecolor=blue, linkcolor=blue, urlcolor=Maroon]{hyperref}
\usepackage{array,tabularx,booktabs,geometry,subfigure,graphicx,longtable}
\usepackage[bf]{caption}
\usepackage{tikz}
\usepackage[all]{xy}
\usetikzlibrary{shapes,arrows}
\tikzstyle{block} = [rectangle, draw, text width=7em, text centered, rounded corners, minimum height=3em]
\usepackage{graphicx,color}
\usepackage{cite}
\usepackage{tikz-cd}

\newtheorem{theorem}{Theorem}[section]

\newtheorem{definition}[theorem]{Definition}

\let\eqref=\relax
\newref{eq}{name={eq.~},Name={Eq.~},names={eqs.~},Names={Eqs.~},rngtxt={-},refcmd=(\ref{#1})}
\newref{f}{name={footnote~},Name={Footnote~},names={footnotes~},Names={Footnotes~}}
\newref{app}{name={appendix~},Name={Appendix~},names={appendixes~},Names={Appendixes~}}
\newref{tab}{name={table~},Name={Table~},names={tables~},Names={Tables~}}
\newref{ch}{name={chapter~},Name={Chapter~},names={chapters~},Names={Chapters~}}
\newref{sec}{name={section~},Name={Section~},names={sections~},Names={Sections~}}
\newref{fig}{name={figure~},Name={Figure~},names={figures~},Names={Figures~}}
\numberwithin{equation}{section}
\newcommand{\eref}[1]{(\ref{#1})}

\newcommand{\eeq}{\end{equation}}
\newcommand{\beq}{\begin{equation}}
\newcommand{\ba}{\begin{array}}
\newcommand{\ea}{\end{array}}

\newcommand{\be}{\begin{equation}}
\newcommand{\ee}{\end{equation}}
\newcommand{\bea}{\begin{equation}\begin{aligned}}	
\newcommand{\eea}{\end{aligned}\end{equation}}		

\newcommand{\iddots}{\mathinner{\mkern2mu\raise1pt\hbox{.}\mkern2mu \raise4pt\hbox{.}\mkern2mu\raise7pt\hbox{.}\mkern1mu}}

\providecommand{\id}{\leavevmode\hbox{\small$\mathrm{1}$\kern-3.8pt\normalsize$\mathrm{1}$}}
\def\fnote#1#2{\begingroup\def\thefootnote{#1}\footnote{#2}
     \addtocounter{footnote}{-1}\endgroup}
\usepackage{blindtext}
\usepackage{scrextend}
\addtokomafont{labelinglabel}{\sffamily}

\begin{document}

\vspace{1cm}

\title{
       \vskip 40pt
       {\huge \bf Extending the Geometry of Heterotic \\
       Spectral Cover Constructions}}

\vspace{2cm}

\author{Lara B. Anderson${}^{1}$, Xin Gao${}^{2}$ and Mohsen Karkheiran${}^{1}$}
\date{}
\maketitle
\begin{center} {\small ${}^1${\it Department of Physics, Robeson Hall, Virginia Tech, Blacksburg, VA 24061, USA}\\
${}^2${\it College of Physics, Sichuan University, Chengdu, 610065, China}

}
\fnote{}{lara.anderson@vt.edu, xingao@scu.edu.cn, mohsenka@vt.edu}
\end{center}

\begin{abstract}
\noindent  
In this work we extend the well-known spectral cover construction first developed by Friedman, Morgan, and Witten to describe more general vector bundles on elliptically fibered Calabi-Yau geometries. In particular, we consider the case in which the Calabi-Yau fibration is not in Weierstrass form, but can rather contain fibral divisors or multiple sections (i.e. a higher rank Mordell-Weil group). In these cases, general vector bundles defined over such Calabi-Yau manifolds cannot be described by ordinary spectral data. To accomplish this we employ well established tools from the mathematics literature of Fourier-Mukai functors. We also generalize existing tools for explicitly computing Fourier-Mukai transforms of stable bundles on elliptic Calabi-Yau manifolds. As an example of these new tools we produce novel examples of chirality changing small instanton transitions. The goal of this work is to provide a geometric formalism that can substantially increase the understood regimes of heterotic/F-theory duality.

\end{abstract}

\thispagestyle{empty}
\setcounter{page}{0}
\newpage

\tableofcontents

\section{Introduction}
Heterotic/F-theory duality has proven to be a robust and useful tool in the determination of F-theory effective physics as well a remarkable window into the structure of the string landscape. The seminal work on F-theory \cite{Morrison:1996na,Morrison:1996pp,Friedman:1997yq} appealed to heterotic theories and ever since, many new developments and tools have been built on, or inspired by, the duality. Despite the important role that this duality has played however, it has remained at some level limited by the geometric assumptions that have been frequently placed on the background geometries in both the heterotic and F-theory compactifications. 

In this work we aim to broaden the consideration of background geometry of manifolds/bundles arising in heterotic compactifications with an aim towards extending the validity and understanding of heterotic/F-theory duality. In particular, we will focus on elliptically fibered Calabi-Yau geometries arising in heterotic theories in the context of the so-called \emph{Fourier Mukai} transforms of vector bundles on elliptically fibered manifolds (see e.g. \cite{Huybrechts} for a review). 

To begin, it should be recalled that compactifications of the $E_8 \times E_8$ heterotic theory on an elliptically fibered Calabi-Yau $n$-fold, 
\beq
\pi_h: X_n  \stackrel{\mathbb E}{\longrightarrow} B_{n-1}\label{hetcy}~~,
\eeq
will lead to the same effective physics as F-theory compactifications on a $K3$-fibered Calabi-Yau $n+1$-fold, 
\beq
\pi_f: Y_{n+1}\stackrel{K3}{\longrightarrow}B_{n-1}~~.\label{f_cy}
\eeq
Here the base manifold $B_{n-1}$ appearing in \eref{hetcy} and \eref{f_cy} is the same K\"ahler manifold (thus inducing a duality fiber by fiber over the base from the 8-dimensional correspondence of \cite{Vafa:1996xn}). Within the heterotic theory, the geometry of the slope stable, holomorphic vector bundle, $\pi: V \to X_n$, must also be taken into account. In particular, to be understood in the context of the fiber-wise duality (induced from $8$-dimensional correspondence), the data of the vector bundle must also be presented ``fiber by fiber" in $X_n$ over the base $B_{n-1}$. 

To this end, the work of Friedman, Morgan and Witten \cite{Friedman:1997yq} introduced the tools of \emph{Fourier-Mukai Transforms} into heterotic theories. In this context, the data of a rank $N$, holomorphic, slope-stable vector bundle $\pi: V \to X$ is presented by its so-called ``spectral data", loosely described as a pair
\beq
(S, {\mathcal L}_S)
\eeq
consisting of an $N$-sheeted cover, $S$, of the base $B_{n-1}$ (the ``spectral cover") and a rank-1 sheaf ${\mathcal L}_S$ over it. Very loosely, this encapsulates the restriction of the bundle to each fiber (given by the $N$ points on the elliptic curve over each point in the base) and the data of a line bundle, ${\mathcal L}_S$ encapsulating the ``twisting" of this decomposition over the manifold. More precisely a Fourier-Mukai transform is a relative integral functor acting on the bounded derived category of coherent sheaves $\Phi: D^b(X) \to D^b(\hat{X})$ (where $\hat{X}$ is the Altman-Kleian compactification of the relative Jacobian of $X$). Let $\mathcal{E} \in D^b(X)$ and define
\begin{eqnarray}
&\begin{tikzcd}
& X \times_B {\hat X} \arrow[dl,"\pi_1 "']\arrow[d,"\rho"]\arrow[dr,"\pi_2 "] &  \\
X & B & \hat{X}
\end{tikzcd}\nonumber\\
& \mathcal E \to \Phi (\mathcal{E}) := R \pi_{2*} (\pi_1^* \mathcal{E}\otimes \mathcal{P}),
\end{eqnarray}
with $X\times_B \hat{X}$ is the fiber product and $\mathcal{P}$ is the ``relative" Poincare sheaf and the so-called ``kernel" of the Fourier-Mukai functor,
\begin{eqnarray}
\mathcal{P} := \mathcal{I}_{\Delta} \otimes \pi_1^* \mathcal{O}_X(\sigma) \otimes \pi_2^*\mathcal{O}_{\hat{X}}(\sigma) \otimes \rho^* K_B^*,
\end{eqnarray}
and where $\mathcal{I}_{\Delta}$ is the ideal sheaf of the relative diagonal divisor,
\begin{eqnarray}
& 0\longrightarrow \mathcal{I}_{\Delta} \longrightarrow \mathcal{O}_{X\times_B \hat{X}} \longrightarrow \delta_* \mathcal{O}_X \longrightarrow 0,\nonumber \\[10pt]
& \delta : X \hookrightarrow X\times_B \hat{X},
\end{eqnarray}
and finally, $K_B$ is the canonical bundle of the base $B$. This functorial/category-theoretic viewpoint will prove a powerful tool as we examine and define the concepts above more carefully in the Sections to come and consider their generalizations.

In the context of heterotic/F-theory duality, a range of possible geometries are possible in the elliptic and $K3$-fibered manifolds appearing in \eref{hetcy} and \eref{f_cy} (with many possible Hodge numbers, Picard groups, etc appearing). However, thanks to the work of Nakayama \cite{nakayama}, the existence of an \emph{elliptic fibration} guarantees the existence of a particular minimal form for the dual CY geometries -- the so-called \emph{Weierstrass form} in which all reducible components of the fiber not intersecting the zero-section have been blown down.

It has been argued \cite{Vafa:1996xn} that from the point of view of F-theory, Weierstrass models are the natural geometric point in which to consider/define the theory. In order to make sense of the axio-dilaton from a type IIB perspective, we require the existence of a section to the elliptic fibration, and for all reducible components of fibers not intersecting this zero section to be blown-down to zero size. This choice provides a unique value of the axio-dilaton for every point in the base geometry. Once it is further demanded that the torus fibration admits a section, it is guaranteed that the Weierstrass models are available and obtainable form the originally chosen geometry via birational morphisms \cite{Deligne}. 

If the F-theory geometry also admits a $K3$-fibration then the choice of Weierstrass form described above also imposes the expected form of the heterotic ellitpically fibered geometry in the stable degeneration limit \cite{Friedman:1997ih,Aspinwall:1997ye,Donagi:2008ca}. As a result, in much of the literature to date, it has simply been assumed that the essential procedure of heterotic/F-theory duality must be to place both CY geometries, $X_{n}$ and $Y_{n+1}$ into Weierstrass form from the start.

However, this Weiestrass-centric procedure overlooks the fact that while the CY manifolds can be naturally transformed into Weierstrass form, the data of a vector bundle in a heterotic theory may \emph{crucially depend on the geometric features that are ``washed out" in Weierstrass form}. In particular, due to a theorem of Shioda, Tate and Wazir \cite{shioda,shioda2,COM:213767}, it is known that the space of divisors of an elliptically fibered CY threefold may be decomposed into the following groups: 

\vspace{2pt}

\noindent 1) Pull-backs, $\pi^*(D_{\alpha})$ of divisors, $D_\alpha$, in the base $B_{n-1}$, \\
\noindent 2) Rational sections to the elliptic fibration (i.e. elements of the Mordell-Weil group of $X_n$), and \\
\noindent 3) So-called ``fibral divisors" corresponding to reducible components of the fiber (i.e. vertical divisors not pulled back from the base). 

\vspace{2pt}

As a result of the above decomposition, it is clear that the topology (i.e. Chern classes), cohomology (i.e. $H^*(X_3,V)$) and stability structure (i.e. stable regions within K\"ahler moduli space) of a stable, holomorphic bundle $V$ on an elliptically fibered manifold can depend on these ``extra" divisors (and elements of $h^{1,1}(X_3)$) which are not present in Weierstrass form. In addition, if $X_n$ contains either a higher rank Mordell-Weil group or fibral divisors, the associated Weierstrass model is singular, leading to natural questions as to how to interpret the data of gauge fields/vector bundles over such spaces. As a result, in the processing of attempting to map the heterotic CY manifold into Weierstrass form, important topological and quasi-topological information -- and its ensuing physical consequences -- could be lost.

It is the goal of this work to investigate Fourier-Mukai transforms of vector bundles over elliptically fibered manifolds \emph{not in Weierstrass form} as a necessary first step in extending heterotic/F-theory duality beyond the form considered in \cite{Friedman:1997yq}.

\vspace{.15in}
The key results of this work include:
\begin{itemize}
\item A generalization of the topological formulae for bundles described by smooth spectral covers to the case of Calabi-Yau threefolds involving fibral divisors and multiple sections (i.e. a higher rank Mordell-Weil group associated to the elliptic fibration).
\item We generalize the available computational tools to \emph{explicitly} construct the Fourier-Mukai transforms of vector bundles on elliptically fibered geometries. That is, given an explicit vector bundle constructed on an elliptic threefold (for example built using the monad construction or as an extension bundle), we provide an algorithm to produce the spectral data (a key ingredient in determining an explicit F-theory dual of a chosen heterotic background). This extends/generalizes important prior work in this area \cite{Bershadsky:1997zv,Lazaroiu,Donagi:2011dv}.
\item We apply the generalized results for spectral cover bundles to the particular application of so-called ``small instanton transitions" in heterotic theories (i.e. M5-brane/fixed plane transitions in the language of heterotic M-theory \cite{Horava:1995qa}.). We find more general transitions possible than those previously cataloged in \cite{Ovrut:2000qi}.
\end{itemize}

The outline of the paper is as follows. In Section \ref{section2} we review the basic tools and key results of Fourier-Mukai transforms and spectral cover bundles in the case of Weierstrass models. We then generalize these results to the case of elliptically fibered manifolds with fibral divisors in Section \ref{section3} and geometries with additional sections to the elliptic fibration in Sections \ref{sectionhol} and \ref{rationalsec}. In Section \ref{section_egs} we provide explicit examples of Fourier-Mukai transforms by beginning with a bundle defined via some explicit construction (e.g. a monad or extension bundle) and then computing its spectral data directly. In Section \ref{section_small_inst} we apply our new results to the problem of chirality changing small instanton transitions. In Section \ref{section_obstruction} we illustrate the distinctions and obstructions that can arise between smooth and singular spectral covers. Finally in Section \ref{section_conclusions} we summarize this work and briefly discuss future directions. The Appendices contain a set of well-known but useful mathematical results on the topics of derived categories and Fourier-Mukai functors. Although the material contained there is well-established in the mathematics literature, it is less commonly used by physicists and we provide a small overview in the hope that readers unfamiliar with these tools might find a brief and self-contained summary of these results useful.

\section{A review of vector bundles over Weierstrass elliptic fibrations and Fourier-Mukai Transforms}\label{section2}
In this section we provide a brief review of some of the necessary existing tools and standard results of Fourier-Mukai transforms arising in elliptically fibered Calabi-Yau geometry. Since the literature on this topic is vast (see for example \cite{Friedman:1997yq, Friedman:1997ih}) and applications \cite{Donagi:1998xe,Donagi:1999gc,Donagi:1999jp,Donagi:1999ez,Donagi:1999fa,Donagi:2000zf,Donagi:2000zs,Donagi:2000fw,Donagi:2004qk,Donagi:2004ia,Curio:1998bva,Curio:2011eu,Andreas:2000sj,Andreas:2006zs,Hayashi:2008ba}, we make no attempt at a comprehensive review, but instead aim for a curated survey of some of the tools that will prove most useful in later Sections. Moreover, we hope that this review is of use in making the present paper somewhat self-contained. However, the reader familiar with this literature could skip straight on to Section \ref{section3}. For more information about the applications of Fourier-Mukai functors in studying the moduli space of stable sheaves over elliptically fibered manifolds, the interested reader is referred to \cite{BBRH}.

\subsection{Irreducible smooth elliptic curve}\label{singlecurve}
To set notation and introduce the necessary tools let us begin by considering the case of $n=1$ in \eref{hetcy}, a one (complex) dimensional Calabi-Yau manifold -- that is $X$ is a smooth elliptic curve, $E$. In the case of a smooth elliptic curve, there is a classical result due to Atiyah \cite{m.atiyah1957} (which can generalized to abelian varieties \cite{BBRH}) 
which states that any (semi)stable coherent sheaf, $\mathcal{E}$, of rank $N$ and degree zero over $E$ is S-equivalent\footnote{For any semistable vector bundle (or torsion free) $V$ with slope $\mu (V)$, there is a filteration -- the Jordan-Holder filteration \cite{Friedman})
of the form $0 = F^0 \subset F^1 \subset \dots \subset F^{k-1} \subset F^k = V$, where $F^i/F^{i-1}$ is stable torsion free with $\mu (F^i/F^{i-1}) = \mu (V)$. Associated with this filteration there is a graded object $gr(V) = \oplus_{i=0}^k F^i/F^{i-1} $, and $V$ and $gr(V)$ are said to be S-equivalent.} to a direct sum of general degree zero line bundles,

\begin{equation}
\mathcal{E} \sim \bigoplus_i \mathcal{L}_i^{\oplus N_i}, \qquad \Sigma_i N_i = N, \quad deg(\mathcal{L}_i)=0.
\end{equation} 

In the context then of the moduli space of semi-stable sheaves on an elliptic curve, one can introduce an integral functor 
\beq
\Phi_{E\rightarrow E}^{\mathcal{P}} : D^b(E) \longrightarrow D^b(E)
\eeq
(note that here $\hat{E}$ the Jacobian of $E$ is simply isomorphic to $E$ and thus we do not make the distinction). This functor admits a canonical kernel, $\mathcal{P}$, the so-called \emph{Poincare sheaf},
\begin{eqnarray}
&\mathcal{P}:= \mathcal{I}_{\Delta} \otimes \pi_1^* \mathcal{O}_E(p_0) \otimes \pi_2^* \mathcal{O}_E(p_0) 
\end{eqnarray}
where $\pi_{1}, \pi_2$ are the projection of $E\times E$ to the first and second factor respectively, $p_0$ is the divisor corresponding to the zero element of the abelian group on the elliptic curves, and $\Delta$ is the diagonal divisor in $E\times E$ (and also $\delta$ is the diagonal morphism). It is not hard to prove that $\mathcal{P}$ satisfies the conditions due to Orlov and Bondal (\cite{BBRH}, see Appendix B)
that guarantee that  $\Phi_{E\rightarrow E}^{\mathcal{P}}$ is indeed a Fourier-Mukai transform (i.e. it is an equivalence of derived categories). 

To illustrate how this specific Fourier-Mukai functor acts on coherent sheaves of degree zero, it is useful to highlight its specific behavior in several explicit cases. To begin, consider the simplest possible case of $\mathcal{E}=\mathcal{O}_E(p-p_0)$, i.e. a generic degree zero line bundle over $E$. Here,

\begin{equation}
\Phi_E^{\mathcal{P}} (\mathcal{O}_E(p-p_0)) = R \pi_{2*} (\pi_1^* \mathcal{O}_E(p-p_0) \otimes \mathcal{P})  \nonumber
\end{equation}

To compute this explicitly, consider the following short exact sequence induced by the morphism $\delta : E\longrightarrow E\times E$,

\begin{equation}
0\longrightarrow \mathcal{P} \longrightarrow \pi_1^* \mathcal{O}_E(p_0) \otimes \pi_2^* \mathcal{O}_E(p_0) \longrightarrow \delta_* \mathcal{O}_E (2 p_0)\longrightarrow 0. 
\end{equation}
Twistin the sequence above with $\mathcal{O}_E(p-p_0)$, and then applying the (left exact) functor $R\pi_*$ to that yields the following long exact sequence (to see the properties of derived functors refer to Appendix A),

\begin{eqnarray}
\begin{tikzcd}[row sep=tiny]
0 \arrow[r] &\Phi^0(\mathcal{O}_E(p-p_0)) \arrow[r] &(R^0 \pi_{2*} \pi_1^* \mathcal{O}_E(p))\otimes \mathcal{O}_E(p_0) \arrow[r] &\mathcal{O}_E(p_0) \otimes \mathcal{O}_E(p)\\
 \arrow[r,hook] &\Phi^1(\mathcal{O}_E(p-p_0)) \arrow[r] &(R^1 \pi_{2*} \pi_1^* \mathcal{O}_E(p))\otimes \mathcal{O}_E(p_0) \arrow[r] &0  .
\end{tikzcd}
\end{eqnarray}

To determine the the FM transform, it is necessary to understand the sheaves appearing in the middle column, and to that end, it is possible to apply the base change formula for flat morphisms,

\begin{eqnarray}
&\begin{tikzcd}
E \times E \arrow[d,"\pi_1 "']\arrow[r,"\pi_2 "] & E \arrow[d,"P"] \\
E\arrow[r,"P"] & p 
\end{tikzcd}\nonumber\\
& R\pi_{2*} \pi_1^* \simeq P^* R P_*,
\end{eqnarray}
where $P$ is just a projection to a point. Therefore, 
\begin{equation}
R \pi_{2*} \pi_1^* \mathcal{O}_E(p) = P^* R\Gamma (E,\mathcal{O}_E(p)) = \mathcal{O}_E.
\end{equation}
Consequently, it follows that $\mathcal{O}_E (p-p_0)$ must be a $WIT_1$, and it is supported\footnote{Note that there is a more intuitive way of getting the same result. The presheaf of the Fourier-Mukai transform of $\mathcal{O}_E (p-p_0)$ over any point $q$ is related to $H^i(E,\mathcal{O}_E(p-q))$, and for $i=0,1$ it is zero unless $p=q$, so naively, both $\Phi^0(\mathcal{O}_E (p-p_0))$ and $\Phi^1(\mathcal{O}_E (p-p_0))$ are some torsion sheaves supported over the point $p$. However, note that since $\mathcal{O}_E (p-p_0)$ is a locally free sheaf, and the projections are flat morphisms, $\Phi^0(\mathcal{O}_E (p-p_0))$ cannot be a torsion sheaf, so only $\Phi^1(\mathcal{O}_E (p-p_0))$ is non-zero, and the only possibility is the skyscraper sheaf $\mathcal{O}_p$.} on $p$,
\begin{eqnarray}
\Phi^{\mathcal{P}}(\mathcal{O}_E (p-p_0)) = \mathcal{O}_p [-1].
\end{eqnarray}
In summary, the Fourier-Mukai transform of any direct sum degree zero line bundles on an elliptic curve, is a direct sum of torsion sheaves supported on the corresponding points of the Jacobian. 

As another simple example, consider the non-trivial extension of two trivial line bundles,
\begin{equation}
0\longrightarrow \mathcal{O}_E \longrightarrow \mathcal{E}_2 \longrightarrow \mathcal{O}_E \longrightarrow 0.
\end{equation}
Applying $\Phi$ on this short exact sequence yields
\begin{eqnarray}
\begin{tikzcd}[row sep=tiny]
0 \arrow[r] &\Phi^0 (\mathcal{O}_E) \arrow[r] &\Phi^0 (\mathcal{E}_2) \arrow[r] & \Phi^0 (\mathcal{O}_E) \\
 \arrow[r,hook] & \Phi^1 (\mathcal{O}_E) \arrow[r] &\Phi^1 (\mathcal{E}_2) \arrow[r]  &\Phi^1 (\mathcal{O}_E) \arrow[r] & 0.
 \end{tikzcd}
\end{eqnarray}
From the previous discussion we have reviewed that $\Phi^{\mathcal{P}} (\mathcal{O}_E) = \mathcal{O}_{p_0}[-1]$, so the first row must be zero (i.e. $\Phi^0 (\mathcal{E}_2)=0$), and

\begin{equation}
0\longrightarrow  \mathcal{O}_{p_0} \longrightarrow \Phi^1 (\mathcal{E}_2) \longrightarrow  \mathcal{O}_{p_0} \longrightarrow  0,
\end{equation}
but this cannot be a non-trivial extension of the torsion sheaves, and one concludes, 
\begin{equation}
\Phi^{\mathcal{P}}(\mathcal{E}_2)=(\mathcal{O}_{p_0}\oplus \mathcal{O}_{p_0}) [-1].
\end{equation}
Note that $\mathcal{E}_2$ is S-equivalent to $\mathcal{O}_E^{\oplus 2}$ but not equal, however, Fourier-Mukai of both of them is the same.  

\subsection{Weierstrass elliptic fibration}

With the results above in hand for a single elliptic curve, they can now be extended fiber-by-fiber for a smooth elliptic fibration. We begin with the simplest case, that of a smooth Weierstrass elliptic fibration $\pi : X \longrightarrow B$. This fibration admits a holomorphic section ${\sigma} : B \rightarrow X$ and every fiber $X_b = \pi^{-1}(b)$ is integral, and generically smooth for $b \in B$. Note that from here onward we will mainly work with smooth Calabi-Yau threefolds and since there exists an isomorphism, $\hat{X}\simeq X$, we will ignore the distinction between $X$ and its relative Jacobian.

In general, on a fibered space, it is possible to define a relative integral functor $\Phi$ in almost the same way it was defined for a trivial fibration (i.e. $ E \times B$, see Appendix \ref{AppendixB} for more information on integral functors).  So for any $\mathcal{E}^{\bullet} \in D^b(X)$ there exists the following:
\begin{eqnarray}
&\begin{tikzcd}
& X \times_B X \arrow[dl,"\pi_1 "']\arrow[d,"\rho"]\arrow[dr,"\pi_2 "] &  \\
X & B & X
\end{tikzcd}\nonumber\\
& \Phi (\mathcal{E}^{\bullet}) := R \pi_{2*} (\pi_1^* \mathcal{E}^{\bullet}\otimes^L \mathcal{K}^{\bullet}),
\end{eqnarray}
with $X\times_B X$ is the fiber product and the kernel is chosen as $\mathcal{K}^{\bullet}\in D^b(X\times_B X)$. In the case at hand,  the kernel is required to be the ``relative" Poincare sheaf,

\begin{eqnarray}
\mathcal{P} := \mathcal{I}_{\Delta} \otimes \pi_1^* \mathcal{O}_X(\sigma) \otimes \pi_2^*\mathcal{O}_X(\sigma) \otimes \rho^* K_B^*,
\end{eqnarray}\label{poinc1}
where $\mathcal{I}_{\Delta}$ is the ideal sheaf of the relative diagonal divisor,
\begin{eqnarray}
& 0\longrightarrow \mathcal{I}_{\Delta} \longrightarrow \mathcal{O}_{X\times_B X} \longrightarrow \delta_* \mathcal{O}_X \longrightarrow 0,\nonumber \\[10pt]
& \delta : X \hookrightarrow X\times_B X,
\end{eqnarray}\label{poinc2}
and $K_B$ is the canonical bundle of the base $B$ (which is chosen to make the restriction $\mathcal{P}|_{\pi_1^*\sigma_1} \simeq \mathcal{O}_X$, and similarly for $\sigma_2$).

From this relative integral functor, it is possible to define ``absolute" integral functor with kernel $j_* \mathcal{P}$, where $j: X\times_B X \hookrightarrow X\times X$ is a closed immersion. Note that $\Phi(\mathcal{E}^{\bullet}) \simeq \Phi^{j_* \mathcal{P}}_{X\rightarrow X} (\mathcal{E}^{\bullet})$ for any $\mathcal{E}^{\bullet}$. It can be proved \cite{BBRH} that this kernel is indeed strongly simple, so the corresponding integral functor is fully faithful. Moreover, since $X$ is a smooth Calabi-Yau manifold, it follows that this integral functor is indeed an equivalence, i.e. a Fourier-Mukai functor. Look at Appendix B, references there. 

It should also be noted that there exist simple formulas for base change compatibility (see Appendix \ref{AppendixB}), and it can be readily verified that the restriction of this Fourier Mukai functor over a generic smooth elliptic fiber is the same as the absolute integral functor that was reviewed briefly in the last Subsection with $p_0$ being the point chosen by the section.   

\subsection{Spectral cover}

It is proved in \cite{Mehta1982} that the restriction of a stable coherent sheaf on a generic fiber is (semi)stable. As we have seen, the relative Fourier-Mukai transform defined in the last subsection, is compatible with base change, and hence its restriction on generic fibers, is the same as the Fourier-Mukai transform on elliptic curves defined in Section \ref{singlecurve}. On the other hand, the Fourier-Mukai transform of a (semi)stable degree zero sheaves of rank $N$ over the elliptic curves is a torsion sheaf of length $N$ (roughly speaking, the support of a torsion sheaf is a set of $N$ points, these points can be infinitesimally close). 

These set of $N$ points over generic fibers define a surface $S \subset X$ and a finite morphism, $\pi_S : S \longrightarrow B$, of degree $N$. This surface $S$  is called a \emph{spectral cover}\footnote{Depending on the choice of gauge group, there are constraints on the position of the points. For example for $SU(n)$ bundles (to which we will restrict our focus in this paper) the sum of these points under the group law of the elliptic curve must be zero. This implies that the spectral cover must be given by a holomorphic function on that torus. For other gauge groups refer to \cite{Friedman:1997yq}, and \cite{Donagi:1997dp}. }, and is the support\footnote{Note that spectral cover can wrap around some elliptic fibers. This is a symptom of the fact that the restriction of the vector bundle over those elliptic fibers is unstable. The restricted Fourier-Mukai transform on these fibers returns non-WIT objects (see Appendix B for definitions), and yet, if $\mathcal{E}$ is a vector bundle, the global Fourier-Mukai still returns a $WIT_1$ object. This is due the flatness of the morphisms and the kernel involved in defining the integral functor.}   
 of $\Phi^1(\mathcal{E})$.
 
On the other hand, the restriction of the torsion sheaf $\Phi^1(\mathcal{E})$ over its support (which is $S$), is a rank one coherent sheaf. This can be seen from the fiberwise treatment (note that $ch_0(\Phi^1(\mathcal{E}))=0$, and $ch_1(\Phi^1(\mathcal{E})) = N =$ Rank $(\mathcal{E})$ when restricted over a generic fiber, since $S$ is actually an $N$-sheeted cover of the base). As a result, the rank of the torsion sheaf over its support must be one (for the cases the support is a non reduced scheme this argument should be modified a little, and it is possible to show that the numerical rank of the spectral sheaf is one, see \cite{BBRH}).  The rank one sheaf $\mathcal{L}:=\Phi^1(\mathcal{E})|_{S}$ is referred to as the \emph{spectral sheaf}, and the doublet $(\mathcal{L},S)$ is called the \emph{spectral data}.

If in addition, if the spectral cover is smooth, the spectral sheaf $\mathcal{L}$ is in fact, a line bundle. In the seminal paper \cite{Friedman:1997yq} some restrictions on the topology of $\mathcal{L}$ are derived, with the assumption that spectral cover $S$ is an integral scheme (reduced and irreducible). We turn to these now, before generalizing them in later sections.

\subsection{Topological data}\label{classic}

A goal of this work is to generalize the results of \cite{Friedman:1997yq} and \cite{Curio:1998vu} for the topology of a vector bundle associated to a smooth spectral cover in the following sections. As a result, it is useful to briefly review the derivation of constraints on the topological data (i.e. the relations between the topology of $\mathcal{L}$ and $ch(\mathcal{E})$). In the following we will assume that the spectral cover is an integral scheme, $\mathcal{E}$ is a $WIT_1$, locally free sheaf (vector bundle) of rank $N$ with vanishing first Chern class, $c_1(\mathcal{E})=0$, and that the Chern character of $\mathcal{E}$ can be written generally as,

\begin{eqnarray}
&&ch(\mathcal{E})= N -c_2(\mathcal{E})+\frac{1}{2} c_3(\mathcal{E}), \nonumber \\[10pt]
&&c_2(\mathcal{E}) = \sigma\eta + \omega [f],\nonumber
\end{eqnarray}
where $\eta$ is the pullback of a base divisor, $[f]$ is the fiber class ($\omega$ is an integer).

We will derive the form of the Chern classes of a smooth spectral cover bundle using a slightly different method than that employed in \cite{Curio:1998vu, Friedman:1997yq}, using tools that are well known in mathematics literature (see for example, \cite{Fulton}) and generalize more readily to the geometries studied in later sections.

Recall that $\Phi(\mathcal{E})=R\pi_{2*} (\pi_1^*\mathcal{E}\otimes \mathcal{P} )$. Thus, we can begin by computing the Chern characters of $\Phi(\mathcal{E})$, using the (singular\footnote{Note that $X\times_B X$ is singular over the discriminant of $X$, even though $X$ is smooth.}) Grothendieck-Riemann-Roch theorem \cite{Fulton} for $\pi_2$:
 
 \begin{equation}\label{GRRclassic}
 ch(\Phi(\mathcal{E})) = \pi_{2*} \left( \pi_1^* ch(\mathcal{E}) ch(\mathcal{P}) td(T_{X/B}) \right),
 \end{equation}
where $td(T_{X/B})$ is the Todd class of the virtual relative tangent bundle of $\pi : X \longrightarrow B$. In addition, it is also necessary to compute the Chern character of the relative Poincare sheaf, and for that, one needs to compute $ch(\mathcal{I}_{\Delta})$. This latter is straightforward to find by applying GRR to the diagonal morphism $\delta$,

\begin{eqnarray}
& 0 \longrightarrow \mathcal{I}_{\Delta} \longrightarrow \mathcal{O}_{X\times_B X} \longrightarrow \delta_* \mathcal{O}_X \longrightarrow 0, \nonumber\\[8pt]
& ch(\mathcal{I}_{\Delta})=1-\delta_* (\frac{1}{td(T_{X/B})}).
\end{eqnarray} 

With these results in place, it remains simply to compute the pullback and push forward of cycles by using the following identities:
\begin{eqnarray}
&&\pi_{2*} \pi_1^* D =0, \quad D \in Div(B), \\[5pt]
&&\pi_{2*} \pi_1^* f =0, \quad f \quad \text{fiber class}, \\[5pt]
&&\pi_{2*} (\pi_1^* c\cdot \delta_* d) = c\cdot d, \quad c,d \in A_{\bullet}(X), \\[5pt]
&&\pi_{2*} (\pi_1^* (\sigma) \cdot b) = b,   \quad b \in A_{\bullet}(B).
\end{eqnarray}  
The first two identities are the result of the fact that if the dimension of the image of a cycle has a lower dimension the corresponding push forward will be zero as a homomorphism between the cycles in the Chow group. The last two follow from the definition of the diagonal morphism and the section (together with projection formula for cycles). 

After putting all of these together, the result is as follows,

\begin{eqnarray}
ch_0 (\Phi(\mathcal{E})) &=& 0 , \label{ch0} \\[5pt]
ch_1 (\Phi(\mathcal{E})) &=& - (N \sigma +\eta) , \label{ch1}\\[5pt]
ch_2 (\Phi(\mathcal{E})) &=& Nn\sigma +\eta) \left(\frac{c_1(B)}{2}\right) + \frac{1}{2} c_3(\mathcal{E}) f, \label{ch2} \\[5pt]
ch_3 (\Phi(\mathcal{E})) &=& -\frac{1}{6}N c_1(B)^2+ \omega . \label{ch3} 
\end{eqnarray}

On the other hand, it should be recalled that $\mathcal{E}$ is $WIT_1$, i.e. $\Phi(\mathcal{E})= i_{S*} \mathcal{L} [-1]$, where $i_{S} : S \hookrightarrow X$, is the closed immersion of $S$ into $X$, and $\mathcal{L}$ is the spectral sheaf (or spectral line bundle in this case). Therefore one can write,
\begin{eqnarray}
ch(\Phi(\mathcal{E})) &=& - ch(i_{S*} \mathcal{L}),  \\
ch(i_{S*} \mathcal{L}) &=& i_{S*} \left( e^{c_1(\mathcal{L})} \frac{1}{T_{X/S}}  \right) \nonumber\\
 &=& {[S]} + {[S]} \cdot \left(c_1(L)-\frac{1}{2} {[S]}\right) + {[S]} \cdot \left(\frac{c_1(\mathcal{L})}{2} -\frac{1}{2}c_1(\mathcal{L}) \cdot {[S]} +\frac{1}{6} {[S]}^2\right).
\end{eqnarray}
where in the second line, the GRR theorem can be applied for the morphism $i_{S*}$, and $T_{X/S}$ is the virtual relative tangent bundle. Importantly, in the third line it is assumed $c_1(\mathcal{L})$ can be written in terms of the divisors of $X$, restricted to $S$, by writing $[S]\cdot c_1(\mathcal{L})$ instead of $i_{S*} \mathcal{L}$ (we'll return to this point in Section \ref{section3}.) 

In summary then, by comparing these two ways of calculating the Chern character of the Fourier-Mukai transform, it is possible to obtain the constraints originally calculated in \cite{Friedman:1997yq,Curio:1998vu}. The first equation (\ref{ch0}) yields simply that $Rank(\Phi^0(\mathcal{E}))-Rank(\Phi^1(\mathcal{E}))=0$, and since we have restricted ourselves to $WIT_1$ sheaves, $\Phi^0(\mathcal{E})=0$ (see Appendix \ref{AppendixB} for definitions), so this means that $Rank(\Phi^1(\mathcal{E}))=0$ i.e. $\Phi^1(\mathcal{E}))$ is a torsion sheaf (which is not surprising). From the first Chern character, the divisor class of the spectral cover can be read (noting the relative minus sign), 
\begin{equation}\label{SClass}
[S]= N \sigma + \eta.
\end{equation}
The next comparison puts non-trivial constraints on $c_1(\mathcal{L})$,
 
\begin{equation}
- [S] \cdot (c_1(\mathcal{L})-\frac{1}{2} {[S]}) =  (N\sigma +\eta) \left(\frac{c_1(B)}{2}\right) + \frac{1}{2} c_3(\mathcal{E}) f.
\end{equation}
Therefore the general form of the first Chern class must be of the form, 

\begin{eqnarray}
c_1(\mathcal{L}) &=& \frac{1}{2} \left( -c_1(B)+{[S]} \right) + \gamma , \\
{[S]}\cdot \gamma &=& -\frac{1}{2} c_3(\mathcal{E}) f.
\end{eqnarray}
The only solution for the second equation above is 

\begin{equation}\label{Gamma}
\gamma = \lambda (N \sigma - \eta + N c_1(B)),
\end{equation}
where $\lambda$ is a constant which can be half integer or integer. So the general solutions for the $c_1(\mathcal{L})$ and $c_3(\mathcal{E})$ are,

\begin{eqnarray}
c_1(\mathcal{L}) &=& \frac{1}{2} \left(- c_1(B)+{[S]} \right) +  \lambda (N \sigma - \eta + n c_1(B)),\label{c1L}  \\
c_3(\mathcal{E}) &=& 2\lambda \eta (\eta-N c_1(B)). \label{c3V}
\end{eqnarray}
where in general, $\lambda$ must satisfy constraints (i.e. be either integer or half integer) in order for $c_1(\mathcal{L})$ to be integeral \cite{Friedman:1997yq}. Note that there is sign difference between (\ref{c1L}), and the similar formula in \cite{Friedman:1997yq}. This arises because either $\mathcal{P}^{\vee}$ or $\mathcal{P}$ may be used as the kernel of the Fourier-Mukai functor. Finally it is possible to obtain $\omega$ from (\ref{ch3}),
\begin{eqnarray}
-\frac{1}{6}N c_1(B)^2+ \omega = -{ [S]} \cdot \left(\frac{c_1(\mathcal{L})}{2} -\frac{1}{2}c_1(\mathcal{L}) \cdot {[S]} +\frac{1}{6} {[S]}^2\right).
\end{eqnarray}
By plugging (\ref{c1L}) and (\ref{SClass}) into this one gets,

\begin{eqnarray}\label{omega}
\omega =-\frac {c_1(B)^2 N^3} {24} + \frac {c_1(B)^2 N} {24} +\frac {1} {8} {c_1(B)}\eta N^2 - \frac {\eta^2 N} {8}- \frac {1} {2} {c_1(B)}\eta\lambda^2 N^2 + \frac {1} {2}\eta^2\lambda^2 N.
\end{eqnarray}
As a result, we arrive finally at the following well-known formulas for the Chern classes of a bundle corresponding to a smooth spectral cover within a Weierstrass CY 3-fold:
\begin{align}
& c_1(\mathcal{E})=0 
\label{fmw1}\\
& c_2(\mathcal{E})=\eta \sigma - \frac{N^3-N}{24}c_1(B_2)^2+\frac{N}{2}\left(\lambda^2-\frac{1}{4}\right)\eta \cdot \left(\eta-Nc_1(B_2) \right) \\
& c_3(\mathcal{E})=2\lambda \sigma \eta \cdot \left(\eta-Nc_1(B_2) \right)
\label{chern_spec}
\end{align}

This is identical with the result of \cite{Friedman:1997yq}. Having reproduced this classic result, we turn in the next section to our first generalization: Fourier-Mukai transforms and spectral cover bundles for elliptically fibered CY 3-folds exhibiting reducible fibers over co-dimension 1 loci in the base (i.e. the 3-folds contain so-called ``fibral" divisors).

\section{Elliptically fibered manifolds with fibral divisors}\label{section3}

In this section we extend the classic results of Section \ref{classic} and consider the Fourier-Mukai transform of a vector bundle over a smooth elliptically fibered Calabi-Yau threefold $\pi : X \longrightarrow B$ with a (holomorphic) section $\sigma$ and so-called \emph{fibral divisors} -- divisors $D_{I}$, $I=1, \ldots m$, which project to a curve in the base $B_2$. In the absence of any additional sections to the elliptic fibration, we have a simple decomposition of the Picard group of $X$ into a) a holomorphic section b) Divisors pulled back from the base, $B$, and c) fibral divisors. Hence, $h^{1,1}(X_3)=1+h^{1,1}(B_2)+m$. Moreover, as a result of the fibral divisors, it is clear that there will be new contributions to the Picard group of $S$, $Pic(S)$ compared to a Weierstrass model. These new geometric integers clearly effect the heterotic theory (and could potentially change the $G_4$ flux present in an F-theory dual geometry). 

Our first effort will be to derive topological formulas for the topology of a bundle over an $X_3$ of the form described above and compare these to the standard case (i.e. \eref{classic} in Section \ref{section2}). We will demonstrate that although the new divisors in $X_3$ do in general effect the topology of possible smooth spectral cover bundles defined over $X_3$, they do not contribute to the chiral index.

In general, the form of the fibral divisors (at co-dimension 1 in $B_2$) will be of the form expected by Kodaira-Tate \cite{kodaira,tate} and a rich array of possibilities is possible. For simplicity, here we will consider the case of $I_n$-type reducible fibers only. It should be noted that even in this simple case, it is clear that the intersection numbers of divisors in $X_3$ and the topology of a spectral cover bundle $\pi: \mathcal{E} \to X_3$ will be more complicated than in the simple case of Weierstrass models considered in Section \ref{section2}. For instance, although some triple intersection numbers of $X_3$ can be simply parameterized in terms of the intersection structure of $B_2$, not all can (see e.g. \cite{Grimm:2013oga} for a list of the triple intersection numbers of an elliptic manifold which are currently known in general). For instance, it is not currently known how to generally parameterize triple intersection numbers involving only fibral divisors in a base-independent way.

Since generic fibers in $X_3$ are still irreducible smooth elliptic curves, we will begin by briefly considering what happens over fibers with ``exceptional curves", taking the case of $I_2$ fibers for simplicity. For more details the interested reader is referred to \cite{caldararu2000derivedThesis,caldararu2000derived,ruiperez2009moduli}.

\subsection{(Semi) stable vector bundles over $I_2$ elliptic curves}

The $I_2$ degeneration of an elliptic fiber is a union of two rational curves $C_1 \cup C_2$ with two intersection points. We assume the section of the elliptic fibration intersects transversely with $C_1$ at a point $p_0$. In general any locally free sheaf $\mathcal{E}$ of rank $N$ over such a reducible fiber can be characterized by its restriction over the components\cite{lopez2005simpson},

\begin{eqnarray}
0\longrightarrow \mathcal{E} \longrightarrow \mathcal{E}_{C_1} \oplus \mathcal{E}_{C_2} \longrightarrow T \longrightarrow 0,
\end{eqnarray} 
where $T$ is a torsion sheaf supported over the intersection points of $I_2$. Now consider a torsion free rank one sheaf $\mathcal{L}$ of degree zero (it is useful to recall that here the notions of degree and rank are defined by the Hilbert polynomial). If $\mathcal{L}$ is strictly semistable, the restrictions $\mathcal{L}_{C_1}$ and $\mathcal{L}_{C_2}$ are $\mathcal{O}_{C_1}(-1)$ and $\mathcal{O}_{C_2}(+1)$ or the other way around. In any case the graded object (defined by the Jordan-Holder filteration) is \cite{lopez2005simpson},
\begin{eqnarray}\label{semistable}
Gr(\mathcal{L})=\mathcal{O}_{C_1}(-1)\oplus \mathcal{O}_{C_2}(-1).
\end{eqnarray} 
On the other hand the graded object of the stable ones are,
\begin{eqnarray}\label{stable}
Gr(\mathcal{L})=\mathcal{O}_{C_1}(p-p_0)\oplus \mathcal{O}_{C_2}.
\end{eqnarray} 
Therefore the graded object of any semi stable bundle over $I_2$ is a direct sum of the cases mentioned above. One can also note that the compactified Jacobian of $I_2$ is a nodal elliptic curve in which all of the semistable line bundles (\ref{semistable}), map to the singular node, and the line bundles map uniquely to the smooth points as in the smooth elliptic curve\cite{lopez2005simpson,ruiperez2009moduli}.

It is proved in \cite{caldararu2000derived,caldararu2000derivedThesis} that the integral functor $\Phi_{I_2\rightarrow I_2}^{\mathcal{P}_0}$ defined by the usual Poincare sheaf $\mathcal{P}_0 = \mathcal{I}_{\Delta}\otimes \pi_1^* \mathcal{O}_{I_2} (p_0)$, satisfies the criteria mentioned in Appendix \ref{AppendixB}, and therefore it is a Fourier-Mukai functor. The action of this functor over the stable line bundles (\ref{stable}) is the same as that defined in Section \ref{section2},
\begin{eqnarray}
\Phi_{I_2\rightarrow I_2}^{\mathcal{P}_0}(\mathcal{L}) = \mathcal{O}_p [-1].
\end{eqnarray}
It remains, then, to compute the other case. Assume $\mathcal{L}=\mathcal{O}_{C_1}(-1)$. As before, by using the exact sequence for $\mathcal{I}_{\Delta}$ and base change formula, one can compute,
\begin{eqnarray}
0\longrightarrow &\Phi_{I_2\rightarrow I_2}^{\mathcal{P}_0 0}(\mathcal{O}_{C_1}(-1)) &\longrightarrow \pi^*\pi_* \mathcal{O}_{C_1} \longrightarrow \mathcal{O}_{C_1} \rightarrow\nonumber \\
\rightarrow &\Phi_{I_2\rightarrow I_2}^{\mathcal{P}_0 1}(\mathcal{O}_{C_1}(-1)) &\longrightarrow \pi^*R^1\pi_* \mathcal{O}_{C_1} \longrightarrow 0,
\end{eqnarray}  
since $\pi^*R\pi_* \mathcal{O}_{C_1}=\mathcal{O}_{I_2}$, and the third map in the first row is surjective, we conclude,
\begin{eqnarray}
\Phi_{I_2\rightarrow I_2}^{\mathcal{P}_0}(\mathcal{O}_{C_1}(-1)) = \mathcal{I}_{C_1}.
\end{eqnarray}
In the same way one finds, 
\begin{eqnarray}
\Phi_{I_2\rightarrow I_2}^{\mathcal{P}_0}(\mathcal{O}_{C_2}(-1)) = \mathcal{O}_{C_2}(-1)[-1].
\end{eqnarray}
Therefore, the Fourier-Mukai transform of a strictly semistable rank one torsion free sheaf \eref{semistable} is, 
\begin{eqnarray}\label{FMofSemistable}
\Phi_{I_2\rightarrow I_2}^{\mathcal{P}_0}(\mathcal{L})=\mathcal{I}_{C_1} \oplus \mathcal{O}_{C_2}(-1)[-1].
\end{eqnarray}
In contrast to the stable line bundles, we see the Fourier-Mukai of \eref{semistable} is non-WIT. However as mentioned before, in the case of elliptic fibration, the Fourier-Mukai transform of a vector bundle can be $WIT_1$ as long as it is stable (and of course flat over the base).

Note that contrary to the case in Section \ref{section2}, the ``Fourier transform" of stable degree zero sheaves over an elliptic fibration $X$ with fibral divisors cannot live in the Jacobian $J(X)$ of $X$. This is because $J(X)$ is indeed a singular variety, and as reviewed in Appendix \ref{AppendixB}, Fourier-Mukai functors are sensitive to singularities, i.e. a singular and a smooth variety cannot be Fourier-Mukai partners. This means if someone tries to ``parameterize" the stable degree zero vector bundles over $X$ by some ``spectral data" in $J(X)$ some important information will be lost. We will return to this in Section \ref{missing}. However, as we will see, it is possible to uniquely ``parameterize" the stable degree zero vector bundle moduli in terms of the resolution of $J(X)$, i.e. $X$ itself.
\subsection{Topological data}

The results of the previous section give us the tools to extend the Fourier-Mukai transform discussed in previous Sections to the singular/reducible fibers present in the case of an elliptic threefold with $I_n$ reducible fibers. In this subsection, the same tools used for Weierstrass models are employed to determine the topology (i.e. Chern classes) of smooth spectral cover bundles on elliptic Calabi-Yau manifolds with fibral divisors. As in Section \ref{section2} we define the an integral functor with Poincare sheaf as the kernel, and as discussed above, it will be Fourier-Mukai again. So it is still possible to use \eref{GRRclassic} to derive some topological constraints.

The only geometric difference within the CY 3-fold is the existence of new fibral divisors $D_I\in Div(X)$ ($I=1,\ldots r$) which in general will not intersect the holomorphic zero section, and in every ``slice" $\pi^* D$ (with $D$ a divisor pulled back from the base) in the intersection $D_I\cdot \pi^*D$ is a $(-2)$-curve\footnote{From now on, in this section, we define the base divisor $D$ as $D:=\frac{1}{\mathcal{S}\cdot \mathcal{S}} \mathcal{S}$, where $\mathcal{S}$ is the ``image" of the fibral divisors in the base. }. 

With these information, the essential non-zero intersections of divisors are,
\begin{eqnarray}
&&\sigma^2 = - c_1 \cdot \sigma, \\
&&\sigma \cdot D_I = 0, \quad for \quad I=1,\dots ,r ,\\
&& h_{\alpha \beta} := \sigma \cdot D_{\alpha} \cdot D_{\beta} \quad \textit{$h_{\alpha \beta}$ is a symmetric, invertible, integral matrix}, \\
&& D_{\alpha} \cdot D_I \cdot D_J = - \mathcal{C}_{IJ}\mathcal{S}\cdot D_{\alpha},\\
&& \textit{For $I_n$:} \quad \mathcal{C}_{I,I}=2,\quad \mathcal{C}_{I,I+1}=-1.
\end{eqnarray}
With the above constraints we can write the second Chern class of the tangent bundle as,
\begin{eqnarray}
c_2(X) = 12\sigma \cdot c_1 + c_2 + 11 c_1^2 + \sum \xi_I D_I.
\end{eqnarray}

Let us turn now to the computation of the topology of a smooth spectral cover bundle. The general form of the Chern character of a bundle $\pi: \mathcal E \to X$ can be expanded as
\begin{eqnarray}
ch(\mathcal{E})= N - (\sigma \eta + \omega f + \sum \zeta_I D_I) + \frac{1}{2} c_3(\mathcal{E})
\end{eqnarray}
where $\zeta$ and $\eta$ are $\mathbb{Q}$-Cartier divisors pulled back from the base $B$. Similar to the Weierstrass case, we can compute the Chern character of $\Phi^{\mathcal{P}}_{X\rightarrow X}(\mathcal{E})$, 
\begin{eqnarray}
ch_0(\Phi(\mathcal{E})) &=& 0 ,\\[3pt]
ch_1(\Phi(\mathcal{E})) &=& -(N \sigma +\eta) , \\[3pt]
ch_2(\Phi(\mathcal{E})) &=& (N \sigma +\eta) \frac{c_1(B)}{2} + \frac{1}{2} c_3(\mathcal{E}) f + \sum \zeta_I D_I \\
ch_3(\Phi(\mathcal{E})) &=& \omega -\frac{1}{6} n c_1(B)^2.
\end{eqnarray}

As explained before, since $\mathcal{E}$ is locally free, $\Phi(\mathcal{E})$ must be $\text{WIT}_1$. If, as in \cite{Friedman:1997yq}, we assume the support of $\Phi^1(\mathcal{E})$, which is the spectral cover $S$ , is a generic integral scheme, then

\begin{eqnarray}
&&\Phi(\mathcal{E})=i_{S*} \mathcal{L}[-1], \\[3pt]
&& i_{S*}: S \hookrightarrow X,
\end{eqnarray} 
where $\mathcal{L}$ must be a line bundle over $S$ as long as $\mathcal{E}$ is given by a smooth spectral cover. After using GRR for the surface $S$, the following results obtained, 
\begin{align}
&{[S]} = n\sigma +\eta ,&\\
&c_1(\mathcal{L}) = \frac{1}{2}(-c_1+[S]) +\gamma + \sum \beta_{iI} e_{iI}, \\
&{[S]}\cdot\gamma =-\frac{1}{2} c_3(\mathcal{E}) f, \\
\end{align}
where $e_{iI}$'s are the fibral (-2)-curves intersecting the spectral cover. $I$ labels the generator of the algebra, $i$ labels the number of the isolated curves (determined by $\eta$). Note that the number of such curves with the spectral cover can be determined by computing the intersection number ${[S]}\cdot D_I^2$ and dividing by ${-2}$. Furthermore, these (-2)-curves intersect as,
\begin{eqnarray}
e_{iI} \cdot e_{jJ}= -\delta_{ij} \mathcal{C}_{IJ}.
\end{eqnarray}

After proceeding as before, we obtain the following solutions,
\begin{eqnarray}
&&\gamma = \lambda(n\sigma-\eta+n c_1(B)), \\[3pt]
&&c_3(V) = 2\lambda \eta (\eta-n c_1(B)), \\[3pt]
&&\omega = \omega_{std} - (-\sum_{i,I} \beta_{iI}^2 + \sum_{i,I}\beta_{iI}\beta_{i,I+1}),\end{eqnarray}
where $\omega_{std}$ is the same as \eref{omega}. However not all parameters $\beta_{iI}$ are free, instead they should satisfy the following equations,
\begin{eqnarray}
\sum_{i}^k \beta_{iI} D\cdot D_I = - \zeta_I \cdot D_I, \quad \text{for each I},
\end{eqnarray}
where $k$ is the number of the ``sets" of $(-2)$-curves inside the spectral cover, 
\begin{eqnarray}
k=  \eta \cdot \mathcal{S}.
\end{eqnarray}

Therefore the only contribution of the $(-2)$-curves will appear in $c_2(\mathcal{E})$ via the correction to (\ref{omega}) (note that similar results were derived in \cite{Donagi:1999ez,Donagi:2000zf}). 

Unlike in the case of Weierstrass models explored in the previous subsection, here it is difficult to write a fully general expression for the Chern classes of $\mathcal{E}$ due to the incomplete knowledge of triple intersection numbers within the CY geometry. In order to make this explicit, we turn to the case of a single fibral divisor here -- that is a CY 3-fold with resolved $SU(2)$ singular fibers.

In this case I=1 and the correction to the second Chern class is of the form,
\begin{eqnarray}
\omega = \omega_{std} + \sum_{i}^{k} \beta_i^2,
\end{eqnarray}

The condition on $\beta_i$ is,
\begin{eqnarray}
(\sum_{i=1}^k \beta_i) \frac{\mathcal{S}}{\mathcal{S}\cdot \mathcal{S}} \cdot D_1 = - \zeta_1 D_1.
\end{eqnarray}
This is equivalent to (by multiplying with $D_1$),
\begin{eqnarray}
\sum_{i=1}^k \beta_i = - \zeta_1 \cdot \mathcal{S}.
\end{eqnarray}

Therefore the correction would be,
\begin{eqnarray}
\omega = \omega_{std} + (\zeta_1 \cdot \mathcal{S} + \sum_{i=2}^k \beta_{i} )^2 + \sum_{i=2}^k \beta_i^2.
\end{eqnarray}
It should be noted that this correction term will contribute to anomaly cancellation in the heterotic theory and to the G-flux in the dual F-theory geometry. We'll return to this point in later sections.
In summary then,
\begin{eqnarray}
&&c_2(\mathcal{E})= \sigma \cdot \eta + \omega_{std}+ (\zeta_1 \cdot \mathcal{S} + \sum_{i=2}^k \beta_{i} )^2 + \sum_{i=2}^k \beta_i^2 + \zeta_1 \cdot D_1, \\
&& c_3(\mathcal{E}) = 2\lambda \eta(\eta-n c_1(B)),
\end{eqnarray}
and $\lambda$ is subject to the same integrality conditions as \cite{Friedman:1997yq}.
\subsection{What is missing in the singular limit}\label{missing}

There is a ``common belief" in the literature that if one need to find the F-theory dual of a perturbative heterotic model on a non Weierstrass elliptically Calabi-Yau with fibral divisors, then one should shrink the exceptional divisors first, and try to find the F-theory dual by working with spectral data in the singular Weierstrass limit. Here we will comment on this from the heterotic string point of view, and explain what will be missed if one uses the naive spectral data in the singular limit.  

As it should be clear by now, the naive spectral data in the singular limit are not in a one to one correspondence  with the bundles in the smooth limit where the exceptional divisors have non zero size i.e. the integral functor is not going to be an equivalence. Hence, if one use the ``singular spectral data" to find the F-theory dual, some information will be lost.

\begin{figure}
\centering
\includegraphics[scale=0.6]{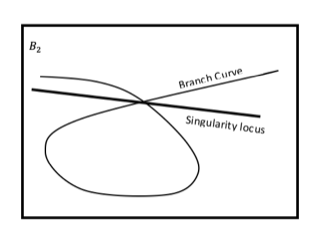}
\caption{Branch locus and singularity locus near the singularity of the spectral cover.}
\label{sing}
\end{figure}

More concretely, as mentioned before, the actual spectral cover in the smooth elliptic fibration will generically wrap around a finite number of $(-2)$-curves, and the spectral sheaf may or may not be dependent on them. So in the blow down limit, the $(-2)$-curves shrink into double point singularities. These singularities are located at the points where the double points of the branch curve intersect with singularity locus of the Weierstrass model i.e. if we look at their image on the base, Fig \eref{sing}, they correspond to the points where the double point singularity of the branch curve hits the singularity locus of the elliptic fibration on the base. On the other hand, locally near these singularities, two sheets of the spectral cover meet each other, and one can use a local model in $\mathbb{C}^3$ as,
\begin{eqnarray}\label{conesing}
S=z^2 - x y=0,
\end{eqnarray} 
where $x,y,z$ are the coordinates of the $\mathbb{C}^3$. Here $S$ is a cone, and can be viewed as the double cover of the $x-y$ plane with branch locus on the lines $x=0$ and $y=0$. The double point singularity is located on the vertex of the coin i.e. $x=y=0$. Now, as it is well known (see for example \cite{Hartshorne} example 6.5.2), the generator of the curve will be a Weil divisor. So instead of the original Cartier $(-2)$-curves on the smooth spectral cover, one gets Weil divisors in the singular limit, and any line bundles on the singular spectral cover will be independent of them.   

Now lets look at the situation the other way around. Suppose we naively choose a generic $n$-sheeted cover of $B_2$ in the singular Weierstrass limit, and a line bundle over that, and use these to find the F-theory dual or study the moduli space of the heterotic string. First of all, for any choice of complex structure of this generic spectral cover, it contains a finite number of double point singularities. To see this, restrict the elliptic fibration over a singular locus where the Weierstrass equation factors as (in the patch $Z=1$),

\begin{eqnarray}
Y^2= (X-b_0)^2 (X-b_1),
\end{eqnarray}
where $b_0$ and $b_1$ are suitable polynomials. In addition a generic $n$-sheeted cover can be written as,
\begin{eqnarray}
S=g_{n-4}(Y) X^2 +g_{n-2}(Y) X+ g_{n}(Y),
\end{eqnarray}
where $g_{n-4}$, $g_{n-2}$ and $g_n$ are polynomial in terms of $Y$ and appropriate local coordinates on base, and the subscripts determine the degree in tems of $Y$ \footnote{For example $Y$ itself is of degree 3 .}. After eliminating $X$ in these to equation we get the following interesting degree $n$ polynomial in terms of $Y$,

\begin{eqnarray}
\left (b_0^2 g_{n-4} + b_0 g_{n-2} + g_{n} \right)^2 \left (b_1^2 g_{n-4} + b_{1}  g_{n-2} + g_n \right)+Y^2 G_{n-2}(Y),
\end{eqnarray}
where $G_{n-2}$ is polynomial in terms $Y$ (of degree $n-2$) and base coordinates which we don't need to know the details. Zeros of this polynomial (with multiplicity) are the points where the $n$-sheeted cover hits the (singular) elliptic curve. Now, note that if  
\begin{eqnarray}\label{BRNCHO1}
b_1^2 g_{n-4} (Y=0) + b_{1}  g_{n-2} (Y=0)+ g_n (Y=0) =0,
\end{eqnarray}
or
\begin{eqnarray}\label{BRNCHO2}
b_0^2 g_{n-4} (Y=0)+ b_0 g_{n-2} (Y=0) + g_{n} (Y=0) =0.
\end{eqnarray}
However from the above equation it is clear that the zeros of \eref{BRNCHO2} are order two, this means the over these points the $n$-sheeted cover is locally like \eref{conesing} (for suitable $x,y,z$) i.e. a double point singularity. The conclusion from the above calculations that we want to emphasize, is that the ubiquitous double point singularities of the $n$-sheeted covers in the singular Weierstrass limit, signals the necessity of working in the blown up limit.

The second problem with ``parameterizing" the vector bundle moduli with the singular data is that since the line bundle in the singular limit doesn't depend on the $(-2)$-curves, the vector bundle that is constructed will not land on some specific components of the moduli space. In particular, physically, at least one consequence of this is missing some new possibilities for the small instanton transitions through exchanging $5$ branes in the Heterotic M-Theory picture. In the context of heterotic/F-theory duality, we expect that $(-2)$-curves inside the spectral cover correspond to new $G_4$-fluxes in the F-theory dual, consistent with the Fourier-Mukai calculations above, and if one considers only the singular spectral cover such possibilities could be missed. 

\section{Non trivial Mordell Weil group with a holomorphic zero section}\label{sectionhol}

In this section we continue our generalization away from Weierstrass elliptic fibrations by considering a Fourier-Mukai transform of vector bundles on elliptically fibered geometries in which the fibration admits more than one section -- that is a higher rank Mordell-Weil group (the group of rational sections to the elliptic fibration \cite{Mordell,Weil}). In the case that the zero section is strictly holomorphic (rather than rational) the definition of the Fourier-Mukai transform introduced in \cite{Friedman:1997yq,Friedman:1997ih} can actually be applied directly. In this case there are also isolated reducible fibers, but as we saw before one can still define a Poincare sheaf, and the corresponding integral functor will be a Fourier-Mukai transform\footnote{Note that if there exists more than one holomorphic section, there is a redundancy in the choice of the ``zero section". The Fourier Mukai functors defined by different choices will be equivalent to each other, and can be written in terms of each other, so we fix the zero section throughout the calculations in this section.}. Therefore, the new Fourier-Mukai functor required for this case is the same as that introduced for fibral divisors in Section \ref{section2}. We defer to later the more generic case of geometries with higher-rank Mordell-Weil group and \emph{only} rational sections (see Section \ref{rationalsec}).

In the case of a holomorphic section and additional (possibly rational) sections, it is clear that the CY 3-fold $X_3$ contains new elements in its Picard group and as a result, their restriction to the spectral cover and $Pic(S)$ will lead to generalizations of the formulas, \eref{classic}, derived in Weierstrass form. We will compute these generalized Chern character formulas directly in the following subsections and independently compare these results to those found in explicit examples in Section \ref{section2} (the latter will be obtained by direct computation of the Fourier-Mukai transforms of a set of simple bundles). 

To set notation, note that we will consider the case of multiple sections to the elliptic fibration and consider the case where the zero section (denoted $\sigma$) is holomorphic. In addition, there are $\sigma_m$ with $m=1, \ldots rk(MW)$ (in general rational) sections present. Here we take $Pic(X)$ of the CY 3-fold to be generated by generated by,

\begin{eqnarray}\label{lncomb}
\sigma\quad \text{the zero section}, \\
S_m = \sigma_m -\sigma -\pi^* \pi_{*} \sigma_m \sigma -c_1(B) , \\
D_\alpha, \quad \alpha=1,\dots h^{1,1}(B). 
\end{eqnarray}
where $S_m$ is the Shioda map of the rational section. Since $\sigma$, there exists a general relation of the form,
\begin{eqnarray}
\sigma \cdot S_p =\sum_{m=1}^{r} D_{m,p} S_m,
\end{eqnarray}
where $D_{m,p}$ are specific divisors in Pic(B). This is because,

\begin{eqnarray}
\sigma^2\cdot S_m = - c_1(B)\cdot \sigma S_m =0 , \\
\sigma\cdot D_b \cdot S_m =0.
\end{eqnarray}

\subsection{Topological data}\label{MWtop}

As in the case of Weierstrass models considered in Section \ref{section2}, we begin by asking what topological formulas can be derived (in as much generality as possible) for a bundle, $\mathcal{E}$ on the manifold above, defined by a \emph{smooth spectral cover}. 

On an elliptic CY 3-fold as described above, the general form of the Chern character of a degree zero vector bundle can be written as

\begin{eqnarray}
ch(\mathcal{E})&=N-(\sigma \eta + \sum_{i=1}^{r} S_i \eta_i+ \omega f )+ \frac{1}{2} c_3(\mathcal{E}), 
\end{eqnarray}
where $N$ is the rank of the bundle, $S_i$ are the image of the Shioda map \cite{shioda,shioda2} of the generators of the Mordell-Weil group, $r$ is the rank of the Mordell Weil group, and $\sigma$ is the zero section we chose. With the help of GRR theorem, one gets the topology of the Fourier-Mukai transform of this bundle. 

\begin{eqnarray}
ch(\Phi(\mathcal{E})) = - (N \sigma + \eta) + (N\sigma+\eta)\frac{c_1(B)}{2} + \sum_{i=1}^{r}S_i \eta_i+ \frac{1}{2} c_3(\mathcal{E}) f + (\omega - \frac{1}{6} N c_1(B)^2).
\end{eqnarray}

Since $\mathcal{E}$ is locally free, it must be $WIT_1$ and $\Phi^1(\mathcal{E})$ will be a torsion sheaf. If  the support of this torsion sheaf is a generic smooth surface, then, 
\begin{eqnarray}
\Phi^1(\mathcal{E})= i_{S*} \mathcal{L},\nonumber
\end{eqnarray}  
where $\mathcal{L}$ is line bundle\footnote{Recall that smoothness of $\mathcal{E}$ implies the smoothness of $\mathcal{L}$ on $S$.}. So by applying GRR to $i_{S}$, topological constraints we are looking for can be obtained,

\begin{eqnarray}
&&[S] = N\sigma+\eta, \\
&&c_1(\mathcal{L}) = \frac{1}{2} (-c_1(B)+[S])+\sum_i^{r} \beta_i S_i +\lambda (N \sigma-\eta +N c_1(B)), \\
&&\sum_{i,j=1}^r S_j (\beta_i ( \eta \delta_{i,j}+N D_{j,i})+\eta_j \delta_{i,j}) =0. = 0, \\
&& c_3(\mathcal{E}) =2\lambda\eta (\eta-N c_1(B)), \\
&&\omega= \omega_{std} - \frac{1}{2} \sum_{m,n,p} \beta_m \beta_n(\eta \delta_{p,m}+N D_{p,n}) S_k S_j.\end{eqnarray}
where the third equation is a constraint on the $\beta_m$'s, and clearly they contribute in Chern characters of $\mathcal{E}$ only through the corrections in $\omega$, and there is not any correction in $c_3(\mathcal{E})$, i.e. the chirality of the effective theory is unchanged.

\subsection{Rank one Mordell-Weil Group}
In this section, we derive explicit correction to the formulas in Section \ref{classic} in the case $rk(MW)=1$. The formulas above can be rewritten as,

\begin{eqnarray}
&&c_1(\mathcal{L}) = \frac{1}{2} (-c_1(B)+[S])+ \beta_1 S_1 +\lambda (N \sigma-\eta +N c_1(B)),\\
&&\sigma\cdot S_1 = D_{11} \cdot S_1, \quad \text{$D_{11}$ is a specific base divisor}, \\
&&\omega = \omega_{std} - \frac{1}{2} \beta^2 (\eta+N D_{11})S_1^2, \\
&&\beta_1 (\eta+N D_{11})S_1 + \eta_1\cdot S_1 =0.
\end{eqnarray}
Note that $\sigma_1$ induces an integral divisor in $S$, so the coefficient of $\sigma_1$ in $c_1(\mathcal{L})$, i.e. $\beta_1$ must be integer, 

\begin{eqnarray}
&&\beta \in \mathbb{Z}.
\end{eqnarray}
This condition fixes $\eta_1$ in terms of $\eta$. More precisely, if one expand $\eta$ and $\eta_1$ in terms of the base divisor, 
\begin{eqnarray}
&&\eta = \eta^{\alpha} D_{\alpha} ,  \\
&&\eta_1 = \eta_1^{\alpha} D_{\alpha}, \\
\end{eqnarray}
then we get the following,
\begin{eqnarray}
&&\eta_1^{\alpha} =- \beta_1 (\eta^{\alpha}+N D_{11}^{\alpha}),
\end{eqnarray}
where  $\beta_1$ is an integer.  Therefore the Chern classes of $\mathcal{E}$ in this case is given by,
\begin{eqnarray}
&&c_2(\mathcal{E})= \sigma \cdot \eta - \beta_1 (\eta+N D_{11}) \cdot S_1 + \left(\omega_{std} -\frac{1}{2} \beta_1^2 (\eta+N D_{11})S_1^2\right) f,\\
&&c_3(\mathcal{E})=2\lambda \eta (\eta-N c_1(B)).
\end{eqnarray}

\section{Non trivial Mordell Weil group with rational generators}\label{rationalsec}

In this section we consider the last piece that will allow us to compute the Fourier-Mukai transform  of vector bundles (or even any coherent sheaf) over any smooth elliptically fibered Calabi Yau variety $\pi : X \longrightarrow B$. In the previous Section we considered the case in which the elliptic threefold with a non-trivial Mordell-Weil Group and (importantly) the zero section was holomorphic. But this is far from the general case, in which all sections to the fibration are birational (i.e. the locus $\sigma=0$ for such a section is birational to $B_2$ rather than equal to it). 

Here we will consider the moduli space of vector bundles over these more general elliptic fibrations. We emphasize again that such information is potentially very important to the study of both the heterotic theory and its F-theory dual. Below, we demonstrate that it is possible in principle for the chirality of the effective theory to change compared to the computation in Weierstrass form. So this case is distinct from those studied in previous Sections.

What makes this situation a little more complicated is that to define a Poincare sheaf one needs a ``true" section (i.e. an inclusion $i_B : B \hookrightarrow X$ such that $ \pi o i_B = id_B$). In that case the section is holomorphic.  The key property is that a holomorphic section intersects \emph{every} fiber at exactly one point. However if the section is rational, this is not satisfied for finitely many fibers containing reducible curves. As a result, the Poincare sheaf will not be a good kernel for the Fourier-Mukai functor. It is not clear at this moment how to deal with this in general, but there are cases which after a flop transition, the zero rational section becomes holomorphic. We restrict ourselves to this in the following, and general case will be studied in a future work.

The key point is that  one can see that derived categories stay ``invariant" under flop transitions (This is the theorem by Bondal and Orlov (see \cite{Huybrechts} Theorem 11.23, and the references therein). So if after a finite number of flop transition one of the sections becomes holomorphic, then it is possible to reduce the problem to one of the cases described before. The disadvantage to this approach is that it is not guaranteed that such flops exists generally.

\subsection{Flop transitions}

Suppose $C \subset X $ be a rational curve in the Calabi-Yau threefold $X$, and $N_{C}X$ is the corresponding normal bundle (obviously, with rank $2$) over $C$. In general one can always blow up $X$ around this curve $p: \tilde{X} \longrightarrow X$, and the corresponding exceptional divisor $E \in Div(\tilde{X})$ will be isomorphic to $\mathbb{P}(N_{C}X)$, which is therefore a $\mathbb{P}^1$ bundle over $C \simeq \mathbb{P}^1$. If $N_{C}X \simeq \mathcal{O}_C(-1)\oplus \mathcal{O}_C(-1)$, then one can show that the exceptional divisor is just a trivial $\mathbb{P}^1$ bundle over another $\mathbb{P}^1$, i.e. $e\simeq \mathbb{P}^1 \times \mathbb{P}^1$ (see e.g. \cite{Huybrechts}). In any case, after blowing $X$ up, one can decide to blow the rational curve $C$ down to get another threefold variety $q:\tilde{X}\longrightarrow X'$. Such geometric birational transformations are called standard flip transitions, and depending on the normal bundle $N_{C}X$, they can change the canonical bundle of the variety. So in general $X'$ is not a Calabi-Yau variety. However in the special case which is described above, $N_{C}X\simeq \mathcal{O}_C(-1)\oplus \mathcal{O}_C(-1)$, the canonical bundle will remain unchanged ($X'$ will be Calabi-Yau), this is called the standard flop transition.

For a general flip transition, the functor $R q_* L p^* : D^b(X)\longrightarrow D^b(X')$, is a fully faithful functor, and its image can be characterized by using the semi-orthogonal decomposition \cite{Huybrechts}. But here we restrict ourselves to the standard flop transitions, and in this case $R q_* L p^*$ will be an equivalence. To be more clear, consider the following diagram,

\begin{equation}
\begin{tikzcd}
 & \tilde{X} \arrow[dl, "p"]\arrow[dr,"q"] & \\
 X & & X'
\end{tikzcd}
\end{equation}
To compute the topological data, we start with a bundle with most general Chern character as before, 
\begin{eqnarray}
ch(\mathcal{E}) = N - (\sigma \eta + \sum_i S_i \eta_i + \omega f)+\frac{1}{2} c_3(\mathcal{E}) , \nonumber
\end{eqnarray}
where $\sigma$ is the rational zero section of $X$, and the Chern character of the object $\mathcal{F}^{\bullet}:=Rp_* q^* \mathcal{E}$ is needed,
\begin{eqnarray}\label{chernofflop}
ch(Rp_* q^* \mathcal{E} ) = p_* (ch(q^*\mathcal{E}) \frac{Td(\tilde{X})}{Td(X')}),
\end{eqnarray}
then, since the zero section is holomorphic in $X'$, we will be able to compute the Chern characters of $\mathcal{F}^{\bullet}$ in $X'$ as in the last section.
To compute \eref{chernofflop}, we can find the relations between the Chern characters of $T\tilde{X}$ and $TX$. To see this, consider the following diagram,

\begin{eqnarray}
\begin{tikzcd}
E :=\mathbb{P}(N_cX) \arrow[d,"g"] \arrow[r, hook, "j"] &\tilde{X} \arrow[d, "p"] \\
c \arrow[r, hook ,"i"] & X
\end{tikzcd}
\end{eqnarray}
One can prove \cite{Fulton} the following short exact functors,
\begin{eqnarray}
\begin{tikzcd}
&&I. \quad 0\arrow[r]& \mathcal{O}_g(-1) \arrow[r]& g^* N_{C}X \arrow[r] & \mathcal{G} \arrow[r] & 0, \\ [5pt]
&&II.\quad 0\arrow[r]& TE \arrow[r]& j^* T \tilde{X} \arrow[r] & \mathcal{O}_g(-1) \arrow[r] & 0,  \\[5pt]
&&III.\quad 0\arrow[r]&  T \tilde{X} \arrow[r] & p^* TX \arrow[r]& j_* \mathcal{G} \arrow[r] & 0, 
\end{tikzcd}
\end{eqnarray}
where the first one is the relative version of the famous Euler sequence\footnote{Therefore, $\mathcal{G}$ is the relative tangent bundle times $\mathcal{O}(-1)$, i.e. $T_g \otimes \mathcal{O}(-1)$}, the second one is the adjunction, and the third sequence is proved by noting that $T\tilde{X}$ and $TX$ are isomorphic almost everywhere (for details see \cite{Fulton}, Chapter 15). In addition if the divisor in the fiber of $g$ is $t$, and we denote the hyperplane in $C$ as $d$ then one can show,
\begin{eqnarray}
&&t^2=2,\quad t\cdot d=1
\end{eqnarray} 
By using these information, and GRR theorem, one can compute the Chern classes of $\tilde{X}$. The result is the following, 
\begin{eqnarray}
&&c_1(\tilde{X}) = - E, \\
&&c_2(\tilde{X}) = p^* c_2(X') + j_* (t - g^* c_1(\mathbb{P}^1)). 
\end{eqnarray}
Using these data we can get the Chern characters in $X'$,
\begin{eqnarray}\label{topolinXp}
ch(Rp_* q^* \mathcal{E}) = N - (\sigma' \eta + \sum_i S_i' \eta_i + \omega f ) +\frac{1}{2} c_3(\mathcal{E}).
\end{eqnarray}
The next part of the calculations will be the same as the previous section, but with intersection numbers in $X'$ not $X$. So is is possible to employ the same formulas in Section \ref{MWtop}, but the intersection formulas are in $X'$ rather than $X$.

\subsubsection{Carrying out the flops explicitly}
The discussion above is somewhat abstract in nature, and as a result, it's helpful to illustrate these geometric transitions in an explicit Calabi-Yau geometry.

We can illustrate the results stated above with the following simple rank 2 bundle defined by extension:
\begin{eqnarray}
\begin{tikzcd}\label{mon_eg1}
0\arrow[r] & \mathcal{O}_X(-\sigma_1+D_b)\arrow[r]& V_2\arrow[r]& \mathcal{O}_X(\sigma_1 - D_b) \arrow[r]& 0~.
\end{tikzcd}
\end{eqnarray} 

For the Calabi-Yau threefold, we will take the anti-canonical hypersurface of the following toric variety,
\begin{eqnarray}
\begin{array}{|cccccccc|c|}\hline
x_1 & x_2 & x_3 & e & u_1 & v_1 & u_2 & v_2 & -K\\
\hline
0 & 0 & 0 & 0 & 1 & 0 & 1 & 0 & 2 \\
0 & 0 & 0 & 0 & 0 & 1 & 1 & 1 & 3 \\
1 & 1 & 1 & 0 & 0 & 2 & 3 & 0 & 8 \\
1 & 1 & 0 & 1 & 0 & 1 & 2 & 0 & 6\\
\hline
\end{array}
\end{eqnarray}
In this manifold, the flop transition described above (which converts a rational section to a holomorphic one) corresponds simply to a different triangulation of the toric polytope. Each triangulation corresponds to a specific Stanley Reisner ideal, 

\begin{eqnarray}
&&\mathcal{I}_{SR1}= \left\lbrace u_1 u_2, x_3 v_1, v_1 v_2, e v_2, x_1 x_2 x_3, x_1 x_2 e\right\rbrace, \\
&&\mathcal{I}_{SR2}= \left\lbrace e u_1, u_1 u_2, v_1 v_2, e v_2, x_1 x_2 x_3, x_1 x_2 e, x_3 v_1 u_2\right\rbrace .
\end{eqnarray} 
In both cases the sections are,
\begin{eqnarray}
&&\sigma_1 = (1,0,0,0), \\
&&\sigma_2 = (-1,1,2,2). 
\end{eqnarray}
However, in the first triangulation, both section are rational, and in particular, $\sigma_1$ wraps around two $(-1)$-curves. After the flop transition, in the second triangulation, the section $\sigma_1$ becomes holomorphic, and the section $\sigma_2$ remains rational, but it wraps around two more $(-1)$-curves (which are the flop transition of the initial ones). 

To fix notation, we denote the sections in the initial geometry as $\sigma_1$, $\sigma_2$ and the sections in the second geometry as $\sigma'_1$, $\sigma'_2$ respectively\footnote{Also note that both geometries contain an exceptional divisor $E$, and $D$ as the hyperplane in the base $\mathbb{P}^2$, which are common to both geometries.}. To find out the corresponding cycles the that $\sigma_1$ wraps around them, we should compute the intersection formulas. So for the first geometry,
\begin{eqnarray}
&& \sigma_1^2 = -c_1(B) \cdot \sigma_1 + \sigma_1 \cdot E, \\
&& \sigma_1\cdot E = D \cdot E -\frac{1}{4} D\cdot S - 2 f, \\
&& \sigma_2^2 = -c_1(B) \cdot \sigma_2 + \frac{19}{4} D \cdot S + D\cdot e + 38 f.
\end{eqnarray}
The corresponding intersection formulas after the flop transition are,

\begin{eqnarray}
&& {\sigma'_1}^2 = -c_1(B) \cdot \sigma'_1, \\
&& \sigma'_1\cdot E = 0, \\
&& {\sigma'_2}^2 = -c_1(B) \cdot \sigma'_2 + 5 D \cdot S'  + 40 f.
\end{eqnarray}
It is clear that the codimension two cycle that is disappearing from the first geometry in the flop transition is,
\begin{eqnarray}
[C] = D \cdot e -\frac{1}{4} D\cdot S - 2 f,
\end{eqnarray}
and the codimension two cycle appearing in the new geometry is,
\begin{eqnarray}
[C']= - D \cdot e + \frac{1}{4} D\cdot S' + 2 f.
\end{eqnarray}
In particular, note that $\sigma'_1\cdot [C'] = + 2$. 

It is also possible to compute the explicit Fourier-Mukai transform of the vector bundle given in \eref{mon_eg1}. The details of such a computation are outlined in Section \ref{section_egs}. Here we simply state the following result to illustrate the general arguments above.

The Chern characters before and after the flop transition are given by
\begin{eqnarray}
&&Ch(V) = 2 - ((2D_b+c_1(B))\sigma_1 + \frac{1}{4} D \cdot S -D\cdot e -(D_b^2-2 f)), \\
&&Ch(Rp_* q^* V_2) = 2 - ((2D_b+c_1(B))\sigma'_1 - D_b^2 + [c']),
\end{eqnarray}
By substituting the formula for the codimension two class $[C']$ we see $V_2$ and $p_* q^* V_2$ have ``the same" Chern class in accordance with the general result of the previous subsection.

\subsection{Comment on the chirality of the effective theory}

Here we want to study the effect of the $(-1)$-curves in the rational zero section in the spectrum of the effective theory. We will fix notation as,
\begin{eqnarray}
\mathcal{F}^{\bullet}:=R p_{*}q^* \mathcal{E}, \nonumber\\
\mathcal{L}^{\bullet}:= \Phi^{\mathcal{P}}_{X'\rightarrow X'}(\mathcal{F}^{\bullet}).
\end{eqnarray}
The goal then is to compute the zero-mode spectrum (i.e. bundle-valued cohomology groups)  of $\mathcal{E}$ in X. Suppose the support of $\mathcal{L}^{\bullet}$ takes the most general form\footnote{The restriction of the support on the generic irreducible fiber is a set of points such that none of them are coincident.}, this task reduces to computation of $R^1 \pi_* \mathcal{E}$ by using Leray spectral sequence. To find this, first notice that inverse functor of $Rq_* Lp^*$ is given by\footnote{Remember that this is Fourier-Mukai functor so it has an inverse.},
\begin{eqnarray}
\mathcal{E} = Rp_*(Lq^*\mathcal{F}^{\bullet} \otimes \mathcal{O}_{\tilde{X}}(e) ). 
\end{eqnarray}
Therefore we get,
\begin{eqnarray}
R \pi_* \mathcal{E} &=& R\pi'_* (\mathcal{F}^{\bullet} \otimes Rq_*\mathcal{O}_{\tilde{X}}(e)) \nonumber\\
&=& R\pi'_* (\mathcal{F}^{\bullet} ),
\end{eqnarray}
where we used $Rq_*\mathcal{O}_{\tilde{X}}(e)=\mathcal{O}_{X'}$. Next, one can use the same techniques as before to compute the $R\pi'_* \mathcal{F}^{\bullet}$ in terms of the ``spectral data" in $X'$,
\begin{eqnarray}
R\pi_* \mathcal{E} =R\pi'_* \mathcal{F}^{\bullet}=R\pi'_*(\mathcal{L}^{\bullet} \otimes \mathcal{O}_{\sigma'}).
\end{eqnarray}
Naively the above result is the same as in the standard cases. But notice that $\mathcal{L}^{\bullet}$ is the Fourier-Mukai transform of a (may be non-WIT or singular) object  $ \mathcal{F}^{\bullet}$ in $D^b(X')$, and it may receive new contributions from the original $(-1)$-curve in $X$.  In the example computed before, the component $[C_2']$ doesn't intersect with the zero section, so the only contribution to the spectrum of the effective theory is through the line bundle over the component $S$. 

\section{Examples of Explicit Fourier-Mukai Transforms}\label{section_egs}

The power of a Fourier-Mukai transform (and its inverse) is that in principle we can move freely between descriptions of stable vector bundles on elliptically fibered manifolds and the spectral data that we have been studying in Sections \ref{section2}, \ref{section3}, and \ref{sectionhol} . In this section we now utilize this potential to explicitly compute FM transforms of stable bundles defined by the monad construction or by extension (see e.g. \cite{Okonek}). Several explicit realizations of this type have been accomplished before in the literature \cite{Bershadsky:1997zv} and we will provide some generalizations. In particular, we will develop general tools that are applicable away from Weierstrass 3-folds.

In these examples, we shall also observe that although we have derived general formulas for bundles defined via \emph{smooth} spectral covers, this proves to be too limited to describe the explicit bundles we consider in the majority of cases. We will return to this point -- namely that there remain important gaps in our description of general points in the moduli space of bundles -- in Section \ref{exampleMonad}.

Beginning with the simplest possible elliptic CY 3-fold geometry -- i.e. Weierstrass form, we will illustrate the ideas that can be generalized to compute the Fourier-Mukai transform of sheaves which are defined by extension sequences or monads. 

\subsection{Bundles defined by extension on Weierstrass CY threefolds}
To illustrate the techniques of taking explicit FM transforms, we begin with the simplest possible extension bundle -- a rank two vector bundle defined by extension of two line bundles:

\begin{eqnarray}\label{rnk2ext}
0 \longrightarrow \mathcal{L}_1 \longrightarrow V_2 \longrightarrow \mathcal{L}_1^{\vee} \longrightarrow 0.
\end{eqnarray}
We require $V_2$ to be stable, and $c_1(V_2)=0$. Note that a necessary (though not sufficient) constraint on the line bundles appearing in this sequence is that  $\mathcal{L}_1$ must not be effective (i.e. have global sections). For such a stable bundle the restriction of $V_2$ over $E_t = \pi^{-1}(t)$ for a generic $t\in B$ is one of the following cases \cite{m.atiyah1957},

\begin{eqnarray}\label{casesofV2}
V_2|_{E_t} &=& \mathcal{O}_{E_t} \oplus \mathcal{O}_{E_t},\nonumber \\
V_2|_{E_t} &=& \mathcal{E}_2 \otimes \mathcal{F}, \quad deg(\mathcal{F})=0, \\
V_2|_{E_t} &=& \mathcal{O}_{E_t}(-p-p_0) \oplus \mathcal{O}_{E_t}(p-p_0). \nonumber
\end{eqnarray} 
In the first case, the support of the Fourier-Mukai sheaf (i.e. spectral cover), will be a non-reduced scheme (supported over the the section $\sigma$). In the second case $\mathcal{E}_2$ is the unique non trivial extension of trivial line bundles, and $\mathcal{F}=\mathcal{O}_{E_t}(p-p_0)$ for some $p$ (here $p_0$ is the point on $E_t$ chosen by the section), but for Weierstrass fibration, $p=p_0$ for generic fibers, and $V_2|_{E_t} = \mathcal{E}_2$. So again the spectral cover will be non-reduced and supported over the zero section. In the final case, the spectral cover can be non-singular. So it is clear that in the majority of cases, we \emph{do not} expect the FM transform of $V_2$ to be in the same component of moduli space as a \emph{smooth} spectral cover of the form described in Section \ref{section2}. We will illustrate this effect with two choices of $\mathcal{L}_1$ below.

Applying the Fourier-Mukai functor to \eref{rnk2ext} produces a long exact sequence involving the FM transform of the line bundles defining $V_2$. Thus, we can compute $\Phi(V_2)$ if we can compute $\Phi(\mathcal{L}_1)$. To begin, the definition of the Poincare sheaf, \eref{poinc1} and \eref{poinc2}, allows us to write the following short exact sequence:
\begin{eqnarray}
0 \longrightarrow \pi_1^* \mathcal{L}_1 \otimes \mathcal{P} &\longrightarrow & \pi_1^* (\mathcal{L}_1\otimes \mathcal{O}_X(\sigma)) \otimes \pi_2^* (\mathcal{O}_X (\sigma) \otimes \pi^* K_B^*) \nonumber\\ 
&\longrightarrow & \delta_* (\mathcal{L}_1 \otimes \mathcal{O}_X (2\sigma) \otimes \pi^* K_B^*) \longrightarrow 0.
\end{eqnarray}
Now, by applying, $R\pi_{2*}$ to the above sequence, we can compute $\Phi(\mathcal{L}_1)$,

\begin{eqnarray}\label{FML}
0 &\longrightarrow &\Phi^0(\mathcal{L}_1) \longrightarrow R^0 \pi_{2*}\pi_1^* (\mathcal{L}_1 \otimes \mathcal{O}_X(\sigma)) \otimes (\mathcal{O}_X (\sigma) \otimes \pi^* K_B^*)\longrightarrow (\mathcal{L}_1 \otimes \mathcal{O}_X (2\sigma) \otimes \pi^* K_B^*) \rightarrow \nonumber \\
&\longrightarrow & \Phi^1(\mathcal{L}_1) \longrightarrow R^1 \pi_{2*}\pi_1^* (\mathcal{L}_1 \otimes \mathcal{O}_X(\sigma)) \otimes (\mathcal{O}_X (\sigma) \otimes \pi^* K_B^*) \longrightarrow 0.
\end{eqnarray}

With these general observations in hand, we will first consider the case where $\mathcal{L}_1=\mathcal{O}_X(D_b)$ with $D_b$ a divisor pulled back from the base, $B_2$. To use \eref{FML}, in this case, $R \pi_{2*} \pi_1^* (\mathcal{L}_1 \otimes \mathcal{O}_X(\sigma))$ must be computed. To accomplish this, we can use the base change formula (see Appendix \ref{AppendixB}), which relates the following push-forwards,
\begin{eqnarray}
&\begin{tikzcd}
X\times_B X \arrow[r, "\pi_1"] \arrow[d,"\pi_2"] & X \arrow[d,"\pi"] \\
X \arrow[r,"\pi"] & B
\end{tikzcd}\nonumber \\
& R\pi_{2*}\pi_1^* \simeq \pi^* R\pi_*
\end{eqnarray}
therefore $R \pi_{2*} \pi_1^* (\mathcal{L}_1 \otimes \mathcal{O}_X(\sigma))= (\pi^* R\pi_* \mathcal{O}_X(\sigma) )\otimes \mathcal{O}_X(D_b)$. On the other hand, by Koszul sequence for the section ($\sigma$) we have,
\begin{eqnarray}
0 \longrightarrow \mathcal{O}_X \longrightarrow \mathcal{O}_X(\sigma) \longrightarrow \mathcal{O}_{\sigma}(K_B) \longrightarrow 0.
\end{eqnarray}
It is well-known for Weierstrass CY elliptic fibration $\pi: X \longrightarrow B$, $R^0\pi_* \mathcal{O}_X = \mathcal{O}_B, R^1\pi_*\mathcal{O}_X= K_B$ (see e.g. \cite{Esole:2017csj}). So the above sequence implies $R\pi_* \mathcal{O}_X(\sigma) = \mathcal{O}_B$ and hence $R \pi_{2*} \pi_1^* (\mathcal{L}_1 \otimes \mathcal{O}_X(\sigma)) = \mathcal{O}_X$. Plugging this into \eref{FML}, we see that this sequence is just Koszul sequence again which is twisted  $\mathcal{O}_X (\sigma) \otimes \pi^* K_B^*$, 
\begin{equation}\label{FML0}
\Phi(\mathcal{L}_1)=\mathcal{O}_{\sigma}(D_b)[-1].
\end{equation}

We can apply this result then to obtain the FM transform of $V_2$ for this chosen line bundle to find
\begin{equation}
0\longrightarrow \mathcal{O}_{\sigma}(D_b) \longrightarrow \Phi^1(V_2) \longrightarrow \mathcal{O}_{\sigma}(-D_b) \longrightarrow 0~.
\end{equation}
In this case by the arguments given above, $\Phi^1(V_2)$ is supported over the section\footnote{It is also possible to have vertical components, depending on the degree of the divisor $D_b$} and its rank (when restricted over the support) is two (the rank is one when restricted to the modified support). As a result, from the arguments above, we do not expect the topology of this bundle to match the formulas given in \eref{GRRclassic} (and indeed they do not though we will not yet make this comparison explicitly).

Let us not contrast this with another (non-generic) choice of line bundle, 
\begin{equation}\label{special_choice}
\mathcal{L}_1 =\mathcal{O}_X(-\sigma+D_b).
\end{equation}
In this case
\begin{eqnarray}
&&\Phi(\mathcal{O}_X(\sigma+D_b)) = \mathcal{O}_X (-\sigma +K_B+D_b),\label{FML+} \\
&&\Phi(\mathcal{O}_X(-\sigma+D_b)) = \mathcal{O}_X (\sigma +D_b)[-1]. \label{FML-}
\end{eqnarray}

For the choice of line bundle in \eref{special_choice}, the extension bundle $V_2$ is defined by a non-trivial element of the following space of extensions:
\begin{equation}\label{Ext1example}
Ext^1(\mathcal{L}_1^{\vee},\mathcal{L}_1)=H^1(X,\mathcal{L}_1^2)=H^0(B,\mathcal{O}_B(2D_b+c_1(B))\oplus \mathcal{O}_B(2D_b-c_1(B))),
\end{equation}
(note that the last equality follows from a Leray spectral sequence on the elliptic threefold (see \eref{Leray}), and $R\pi_* \mathcal{O}_X(-2\sigma)=K_b\oplus K_b^{-1}$. As a brief aside, we remark here that the form of this space of extensions gives us some information about the form of the possible FM dual spectral cover.

It is clear from the expression above that if $2D_b+c_1(B)$ is not effective, then there exists no non-trivial extension, and the vector bundle is simply a direct sum $\mathcal{L}_1 \oplus \mathcal{L}_1^{\vee}$ (and therefore not strictly stable). If $2D_b+c_1(B)=0$ there is only one non-zero extension. On the other hand, if the degree of $D_b$ is large enough to make $2D_b-c_1(B)$ effective then for any generic choice of extension there are $(2d_b+c_1(B))\cdot (2D_b-c_1(B))$ isolated curves which the spectral cover must wrap.

Returning to our primary goal of computing the FM transform of $V_2$, it can be observed that there is enough information in \eref{FML+} and \eref{FML-} to compute $\Phi(V_2)$ explicitly.
\begin{eqnarray}
0\longrightarrow \Phi^0(V_2) \longrightarrow \mathcal{O}_X(-\sigma+K_B-D_b) \xrightarrow{\text{F}} \mathcal{O}_X(\sigma+D_b) \longrightarrow \Phi^1(V_2)\longrightarrow 0.
\end{eqnarray} 
By fully faithfulness of Fourier-Mukai functor, one can show $F\in Ext^0(\mathcal{O}_X(-\sigma+K_B-D_b),\mathcal{O}_X(\sigma+D_b))\simeq Ext^1(\mathcal{L}_1^{\vee},\mathcal{L}_1)$. Therefore it is necessary $2D_b-c_1(B)$ be effective to have a non zero $F$, and $\Phi^0(V_2)=0$ (and hence stability of $V_2$). Assuming that this is satisfied, we can find the Fourier-Mukai transform of $V_2$ as
\begin{eqnarray}\label{FMV2Example}
\Phi(V_2)=\mathcal{O}_{2\sigma+2D_b-K_B}(\sigma+D_b).
\end{eqnarray}

At last we are in a position to compute the topological data, and directly compare the bundle constructed here with what would be expected from the formulas derived in \cite{Friedman:1997yq,Curio:1998vu} and reviewed in Section \ref{section2}. The Chern character of $V_2$ is,

\begin{eqnarray}
ch(V_2)=2-(\sigma (2D_b+c_1(B))+D_b^2).
\end{eqnarray}
Therefore from \eref{SClass}, the divisor class of spectral cover must be 
\begin{equation}
[S]=2\sigma+2D_b+c_1(B).
\end{equation}
This is the same as the divisor class of the support of the torsion sheaf in \eref{FMV2Example}, In addition, since we require $[S]$ to be the divisor class of a algebraic surface it must be the case that $2D_b+c_1(B)$ is effective. This was exactly the requirement for the non trivial extension discussed above. 

For this example, the general algebraic formula for $S$ takes the form
\begin{eqnarray}
S &=& f_1 x + f_2 z^2, \\
div(f_1) &=& 2D_b-c_1(B), \nonumber \\
div(f_2) &=& 2D_b+c_1(B). \nonumber
\end{eqnarray}
So we see if $2D_b+c_1(B)$ is effective, but $2D_b-c_1(B)$ is not effective, then the coefficient $f_1$ vanishes, and the locus $f_2=0$ is the position of the vertical components mentioned above. Moreover, when $2D_b-c_1(B)$ is effective then the position of those vertical fibers is given by the points where $f_1=f_2=0$, again as discussed before. Comparing this with the sequence before \eref{FMV2Example}, we see the map $F$ is indeed given by $S$, and therefore $S$ uniquely determines an element in the extension group.

 Now from the equation (\ref{c1L}), $c_1(\mathcal{L})=\sigma+D_b+\lambda(2\sigma+2D_b+c_1(B))$. This is compatible with (\ref{FMV2Example}) if we choose $\lambda=0$. With $\lambda=0$ and $N=2$, the equation \eref{omega} produces
\begin{equation}
\omega=D_b^2,
\end{equation}
and also from \eref{c3V} it follows that $c_3(v_2)=0$, in agreement with the Chern character computed directly above. Also note that the divisor class of the matter curve must be $\sigma \cdot [S]=2D_b-c_1(B)$ \cite{Friedman:1997yq}. So the FM transform of this vector bundle is indeed a smooth spectral cover and agrees with the topological formulas found in \cite{Friedman:1997yq,Curio:1998vu} as expected.

\subsection{FM Transforms of Monad Bundles over Weierstrass 3-folds}\label{exampleMonad}
In the following section we will provide an explicit construction of the spectral data a bundle defined via a monad. This construction is somewhat lengthy, but is useful to present in detail to demonstrate that FM transforms can be explicitly constructed for bundles that appear frequently in the heterotic literature.

Over a Weierstrass CY 3-fold of the form studied in Section \ref{section2} consider a bundle defined as a so-called ``monad"  (i.e. as the kernel of a morphism between two sums of line bundles over $X_3$):
\begin{eqnarray}\label{mon_def}
0\longrightarrow V\longrightarrow \oplus_{i=1}^l \mathcal{O}_X(n_i \sigma + D_i) \xrightarrow{\text{F}} \oplus_{j=1}^{k} \mathcal{O}_X(m_j \sigma + D_j) \longrightarrow 0,
\end{eqnarray}
where $Rank(V) = N =l-k$, and the divisors $D_i$ are pulled back from the base, $B_2$. To compute the Fourier-Mukai transform $V$ we will see that it is necessary to begin with the transform of line bundles of the form $\mathcal{O}_X(n_i \sigma + D_i)$, as well as the morphism $\Phi(F)$. With that information, we can compute $\Phi(V)$.  We should point out that for the geometry in question, none of the $n_i$'s nor $m_j$'s are allowed to be negative. This is necessary for stability of the bundle\footnote{Actually if we naively compute the Fourier-Mukai of such sheaves (with some $n_i$'s being negative), the result is either non-$WIT_1$ or $\Phi^1(V)$ is not a torsion sheaf. But we know $V$ is stable if and only if it is $WIT_1$ respect to $\Phi$, and $\Phi^1(V)$ is a torsion sheaf. In practice, this is a way to check the stability of a degree zero vector bundle over elliptically fibered manifolds.}. Upon applying the FM functor to \eref{mon_def}, we get a sequence of the following form,
\begin{eqnarray}
\begin{tikzcd}
0\arrow[r] & \Phi^0(V)\arrow[r] & \oplus_{i=1}^{'l} \Phi^0(\mathcal{O}_X(n_i \sigma + D_i)) \arrow[r,"\Phi(F_0)"] &\oplus_{j=1}^{'k} \Phi^0(\mathcal{O}_X(m_j \sigma + D_j))&  \\
 \arrow[r, hook]&\Phi^1 (V) \arrow[r] & \oplus_{i=1}^{''l}\Phi^1( \mathcal{O}_X(n_i \sigma + D_i)) \arrow[r]& \oplus_{j=1}^{''k} \Phi^1(\mathcal{O}_X(m_j \sigma + D_j)) \arrow[r] & 0.
\end{tikzcd}
\end{eqnarray}
In the diagram above we employ the sign $\oplus^{'}$ to refer to the direct sum over the line bundles with positive definite relative degree, and use $\oplus^{''}$ to mean the direct sum over the line bundles with with relative degree zero  (i.e. pull back of line bundles in the base). So to compute the Fourier-Mukai transform of $V$ we need to compute the Fourier-Mukai transform of the line bundles in \eref{mon_def}. To do this, one can simply use the defining sequence of the diagonal divisor in Section \ref{section2}. Combining this with the sequence above, give the following diagram,
{\scriptsize
\begin{eqnarray}\label{FMofMonad}
\begin{tikzcd}[transform shape, nodes={scale=0.92}]
 & & 0\arrow[d] & 0 \arrow[d] & & & & \\
& \dots \arrow[r]& \oplus_{i=1}^{'l} \Phi^0(\mathcal{O}_X(n_i \sigma + D_i)) \arrow[r,"\Phi(F_0)"]\arrow[d]  &\oplus_{j=1}^{'k} \Phi^0(\mathcal{O}_X(m_j \sigma + D_j)) \arrow[d]\arrow[r] & \dots & & &   \\
0\arrow[r] & K_1 \arrow[r]\arrow[d] & \mathcal{A} \otimes \mathcal{O}_X(\sigma+c_1(B))\arrow[d,"ev"] \arrow[r,"F_0"] & \mathcal{N} \otimes \mathcal{O}_X(\sigma+c_1(B)) \arrow[d,"ev"]\arrow[r] & Q_1 \arrow[d] \arrow[r] & 0 & &\\
0\arrow[r]& K_2 \arrow[r]& \oplus_{i=1}^{'l} \mathcal{O}_X((n_i+1) \sigma + D_i) \otimes \mathcal{O}_X(\sigma+c_1(B)) \arrow[d]\arrow[r,"F_0"] & \oplus_{j=1}^{'k} \mathcal{O}_X((m_j+1) \sigma + D_j) \otimes \mathcal{O}_X(\sigma+c_1(B)) \arrow[d] \arrow[r] & Q_2 \arrow[r] & 0 & &\\
 & & 0 & 0 & & & &\\
 & & & & & & &
\end{tikzcd}
\raisetag{30pt}
\end{eqnarray}}
\\
 Each column in the diagram is defines the Fourier-Mukai transform of the (direct sum of) line bundles by means of the resolution of the Poincare sheaf. Therefore in the second row $\mathcal{A}$ and $\mathcal{N}$ are the sheaves generated by the ``fiberwise" global sections of the sheaves $\oplus^{'} \mathcal{O}_X((n_j+1) \sigma + D_j)$ and $\oplus^{'} \mathcal{O}_X((m_j+1) \sigma + D_j)$, respectively. The evaluation maps simply takes the global section, and evaluates the sheaf at each point. Finally, the map $F_0$ is simply the map induced by the monad map $F$ itself (from \eref{mon_def}) on the line bundles with positive definite relative degree (which also acts on the ``fiberwise" global sections too).

The most important parts of this diagram are the induced maps between the kernels and co-kernels, $K_1$, $Q_1$ and $K_2$, $Q_2$, respectively. The kernel and co-kernel of these maps give a rather explicit presentation of the spectral data, so we will give them specific names,
\begin{eqnarray}
0\longrightarrow \bar{\mathcal{L}} \longrightarrow K_1 \longrightarrow K_2 \longrightarrow \mathcal{L} \longrightarrow 0, \label{LLbar} \\
0\longrightarrow \mathcal{M} \longrightarrow Q_1 \longrightarrow Q_2 \longrightarrow 0, \label{M}
\end{eqnarray}
(note that the final map in the second line above must be surjective, otherwise it will be in contradiction with the commutativity of the middle two columns in \eref{FMofMonad}).  

Now, by careful diagram chasing, one can prove that the Fourier-Mukai transform of $V$ can be given by the following (more consise) diagram, 
\begin{eqnarray}
\begin{tikzcd}
 & 0\arrow[d] & & & & \\
 & \mathcal{L}\arrow[d] & & & & \\
0\arrow[r] & \mathcal{J}\arrow[d] \arrow[r] & \Phi^1 (V) \arrow[r] & \oplus_{i=1}^{''l}\Phi^1( \mathcal{O}_X(n_i \sigma + D_i)) \arrow[r] & \oplus_{j=1}^{''k} \Phi^1(\mathcal{O}_X(m_j \sigma + D_j)) \arrow[r] & 0 \\
& \mathcal{M}\arrow[d] & & & & \\
& 0 & & & &
\end{tikzcd} \hspace{10pt}
\end{eqnarray}
This construction is similar in spirit to the spectral data derived for monads in \cite{Donagi:2011dv} and we will return to this in Section \ref{comparisonwDW}.

To make this abstract formalism more concrete, it is helpful to consider an explicit example. Let us take $X_3$ to be a Weierstrass elliptically fibered threefold over $\mathbb{P}^2$, realized as a hypersurface in a toric variety, given by the following ``charge data" (i.e. in GLSM notation):

\begin{center}
\begin{tabular}{cccccc|c}
y & x & z & $x_0$ & $x_1$ & $x_2$ & p  \\
3 & 2 & 1 & 0 & 0 & 0 & 6 \\
9 & 6 & 0 & 1 & 1 & 1 & 18
\end{tabular}
\end{center}

Here the holomorphic zero section is determined by the divisor $z=0$. As an explicit monad bundle over this manifold, consider the following short exact sequence:

\begin{eqnarray}\label{monadEX}
0\longrightarrow V \longrightarrow \mathcal{O}_X(2,3)\oplus \mathcal{O}_X(1,6)\oplus \mathcal{O}_X(0,1)^{\oplus 3}\xrightarrow{F} \mathcal{O}_X(3,12) \longrightarrow 0.
\end{eqnarray}

We first need to find the Fourier-Mukai of the line bundles. This can be done using the tools outlined in before and we simply summarize the results here:
\begin{eqnarray}
&& \Phi(\mathcal{O}_X(D)) = \mathcal{O}_{\sigma}(K_B + D)[-1],\label{FMofTrivial} \\
\hspace{10pt}\nonumber\\
&& 0\longrightarrow \Phi^0(\mathcal{O}_X(2\sigma - K_B)) \longrightarrow \mathcal{O}_X(\sigma -2 K_B) \oplus \mathcal{O}_X(\sigma) \oplus \mathcal{O}_X(\sigma + K_B) \xrightarrow{ev} \mathcal{O}_X(4\sigma -2 K_B) \longrightarrow 0, \nonumber\label{FMof22} \\ 
\hspace{10pt}\\
&& 0\longleftarrow \Phi^0(\mathcal{O}_X(\sigma - 2 K_B)) \longrightarrow \mathcal{O}_X(\sigma-3K_B) \oplus \mathcal{O}_X (\sigma - K_B) \longrightarrow \mathcal{O}_X(3\sigma -3 K_B) \longrightarrow 0,\nonumber\label{FMof16} \\
\hspace{10pt}\\ 
 && 0\longrightarrow \Phi^0(\mathcal{O}_X(3\sigma - 4 K_B)) \longrightarrow \mathcal{O}_X(\sigma -5 K_B) \oplus \dots \oplus \mathcal{O}_X(\sigma - K_B) \xrightarrow{ev} \mathcal{O}_X(5\sigma -5 K_B) \longrightarrow 0,\nonumber\label{FMof312} \\ 
 \hspace{10pt}
\end{eqnarray}
where the middle bundles in the each of the short exact sequences above are the ``fiberwise" global section of the line bundles in \eref{mon_def} denoted as $\mathcal{A}$ and $\mathcal{N}$ (twisted with $\mathcal{O}(\sigma+c_1(B))$). With this we have determined the columns of (\ref{FMofMonad}). By explicitly performing the fiber restrictions it can also be verified that

\begin{eqnarray}
&&\oplus_{i=1}^{''l}\Phi^1( \mathcal{O}_X(n_i \sigma + D_i)) = \mathcal{O}_{\sigma}(-2)^{\oplus 3}, \nonumber\\
&&\oplus_{i=1}^{''l}\Phi^1( \mathcal{O}_X(m_i \sigma + D_i)) = 0, \nonumber
\end{eqnarray}
and the map $F_0$ is a ``part" of the monad map $F$,
\begin{eqnarray}
&&\begin{tikzcd}
\mathcal{O}_X(2,3)\oplus \mathcal{O}_X (1,6) \arrow[r,"F_0"] &\mathcal{O}_X(3,12)
\end{tikzcd} ,\nonumber \\
&& F_0 = \left( \begin{array}{c}
z f_9 \\
x+f_6 z^2
\end{array}\right) . \label{F0}
\end{eqnarray}  
Obviously $F_0$ is singular on $\left\lbrace f_9=0 \right\rbrace \cap \left\lbrace x+f_6 z^2 = 0\right\rbrace$. 

The final task will be determining the explicit kernels and co-kernels: $K_1$, $K_2$, $Q_1$ and $Q_2$. This is local question, so we can assume we are in a affine patch with $y\ne 0$ and $x_1\ne 0$ for example. Then it is not too hard to show that free part of $K_1$ is generated by  
\begin{eqnarray}
K_1 \sim \alpha z \left( \begin{array}{c}
x+f_6 z^2 \\
-f_9 z
\end{array} \right) . 
\end{eqnarray}
Naively, it may look like that over $f_9=0$, the kernel $K_1$ jumps, but this is at the presheaf level, one can actually show that 
\begin{eqnarray}
K_1\simeq \pi^*\mathcal{O}_{P^2}(-3).
\end{eqnarray}
Similarly, one can compute the $K_2$,
\begin{eqnarray}
&& K_2 =  \left( \begin{array}{c}
(x+f_6 z^2) \frac{1}{l_{3,3}}\\
-z f_9 \frac{1}{l_{2,6}}
\end{array}\right)
\end{eqnarray} 
Where $\frac{1}{l_{3,3}}$ and $ \frac{1}{l_{2,6}}$ are the local generators of the line bundles $\mathcal{O}_X(3,3)$ and $\mathcal{O}_X(2,6)$. By checking the degrees, $K_2$ is fixed to be the line bundle $\mathcal{O}_X(1,-3)$. Again naively it might appear that $K_2$ jumps over $\left\lbrace f_9=0 \right\rbrace \cap \left\lbrace x+f_6 z^2 = 0\right\rbrace$, but this is at the presheaf level as before, and $K_2$ is indeed free. 

With this information in hand, we can determine $\mathcal{L}$ and $\bar{\mathcal{L}}$ in (\ref{FMofMonad}),
\begin{eqnarray}
\begin{tikzcd}
0\arrow[r] &\bar{\mathcal{L}} \arrow[r] &\mathcal{O}_X(0,-3)\otimes\mathcal{O}_X(1,3)  \arrow[r,"\Psi_0"]  &\mathcal{O}_X(1,-3)\otimes \mathcal{O}_X(1,3) \arrow[r] &\mathcal{L} \arrow[r] & 0,
\end{tikzcd}
\end{eqnarray}
By computing the induced map $\Psi_0$, one finds
\begin{eqnarray}
&&\bar{\mathcal{L}}=0, \\
&&\mathcal{L}= \mathcal{O}_{\sigma}(-6).
\end{eqnarray}

As the next step, it remains to determine $Q_1$ and $Q_2$. For the former, one should note that the morphism on the ``fiberwise" global sections i.e. $\mathcal{A} \xrightarrow{F_0}\mathcal{N}$ is generically rank 4, so it is surjective unless $f_9=0$. Over this locus, we obtain the following ``defining" sequence for $Q_1$,

\begin{eqnarray}
0 &\longrightarrow& (\mathcal{O}_X \oplus \mathcal{O}_X(0,6))|_{f_9} \otimes \mathcal{O}_X(1,3) \rightarrow \dots\nonumber \\
&\longrightarrow& (\mathcal{O}_X\oplus \mathcal{O}_X(0,3)\oplus \mathcal{O}_X(0,6)\oplus \mathcal{O}_X(0,12))|_{f_9} \otimes \mathcal{O}_X(1,3) \longrightarrow Q_1\longrightarrow 0. \hspace{10pt}
\end{eqnarray} 
Which turns out to be,
\begin{eqnarray}
Q_1\simeq  \left(\mathcal{O}_X(0,12)\oplus\mathcal{O}_X(0,3)\right)_{f_9=0} \otimes\mathcal{O}_X(1,3).
\end{eqnarray}
On the other hand, $Q_2$ can be identified easily with $\mathcal{O}_X(4,12)|_{\lbrace f_9=0 \rbrace \cap \lbrace x+f_6z^2 =0 \rbrace} \otimes \mathcal{O}_X(1,3)$. So $\mathcal{M}$ will be given by,

\begin{eqnarray}
0\longrightarrow \mathcal{M} \longrightarrow \left(\mathcal{O}_X(0,12)\oplus\mathcal{O}_X(0,3)\right)_{f_9=0} \otimes\mathcal{O}_X(1,3) \longrightarrow \mathcal{O}_X(4,12)|_{\lbrace f_9=0 \rbrace \cap \lbrace x+f_6z^2 =0 \rbrace} \otimes \mathcal{O}_X(1,3) \longrightarrow 0.\nonumber\\
\hspace{5pt}
\end{eqnarray}

Therefore, $\mathcal{M}$ will be a torsion sheaf supported on $f_9=0$ with rank 2 when restricted on the support. So $\mathcal{J}$ in (\ref{FMofMonad}) can be given explicitly as,

\begin{eqnarray}
0\longrightarrow \mathcal{O}_{\sigma}(-6) \longrightarrow \mathcal{J} \longrightarrow \mathcal{M} \longrightarrow 0,
\end{eqnarray} 
and we can see the support of $\mathcal{J}$ is in the divisor class $\sigma+18 D$ where the $18D$ is the support of the sheaf $\mathcal{M}$. Finally the support of the $\Phi^1(V)$, i.e. the spectral cover, is in the class
\begin{eqnarray}
[S] = 4\sigma +18 D.
\end{eqnarray} 
Explicitly we find that the spectral cover is reducible and non-reduced and given by the algebraic expression
\beq
S: (f_9)^2z^4=0
\eeq

With this spectral data in hand we are now in a position to compare to the well-known results for the topology of smooth spectral cover bundles derived in Section \ref{section2}. Before beginning this computation we must first observe that from the definition of the monad in \eref{monadEX}, the Chern class of $V$ is given by,
\begin{eqnarray}
c(V) = 1+ 18 \sigma D+ 48 f - 162 w,
\end{eqnarray}
where $f$ is the fiber class, and $w$ is the class of a point. Now if one compares this to the topological constraints reviewed in \eref{GRRclassic}, it follows that $\eta = 18D$ and hence

\begin{eqnarray}
[S]= 4\sigma + 18 D, \\
c_3(V) = 2 \lambda \eta(\eta-4c_1(B)).
\end{eqnarray}
The first one is always true whether or not the spectral cover is degenerate or what spectral sheaf we choose, so it is not surprising to get a correct answer. The second equation however implies that $\lambda= -\frac{3}{4}$.  If we then insert this value into the formula for the $c_2(V)$ given in \eref{GRRclassic}, it yields 
\beq
c_2(V)_{expected}=18\sigma D +45 f
\eeq 
which is obviously wrong. This discrepancy has arisen because the chosen monad bundle manifestly does not correspond to a smooth spectral cover (and must correspond to a different component of the moduli space of bundles over $X_3$).

\subsubsection{A comparison to existing techniques for FM transforms of monad bundles}\label{comparisonwDW}
It should be noted that several existing papers in the literature \cite{Bershadsky:1997zv,Donagi:2011dv} have laid out useful algorithms for explicitly computing the FM transforms of monad bundles of the form
\begin{eqnarray}
\begin{tikzcd}
0\arrow[r] & V \arrow[r] &\mathcal{F} \arrow[r,"F"] &\mathcal{N} \arrow[r]& 0,
\end{tikzcd}
\end{eqnarray}
where $\mathcal{F}$ and $\mathcal{N}$ are direct sum of line bundles as mentioned before. 

In particular, \cite{Bershadsky:1997zv} utilizes the simple and useful observation that the ``fiberwise" global sections of the twisted vector bundle $V\otimes \mathcal{O}_X(\sigma)$ contain information about the spectral cover. Specifically, the zeros of these sections along the fiber are coincident with the points where the spectral cover intersects the fibers. So one can consider the kernel of the map $F$ in the following sequence,
\begin{eqnarray}
\begin{tikzcd}
0\arrow[r]& \pi^*\pi_*(V\otimes \mathcal{O}_X(\sigma)) \arrow[r]& \pi^*\pi_*(\mathcal{F}\otimes \mathcal{O}_X(\sigma)) \arrow[r,"F"] &\pi^*\pi_*(\mathcal{N}\otimes \mathcal{O}_X(\sigma)) \arrow[r]& 0,
\end{tikzcd}
\end{eqnarray}
where the morphism $\pi$ is the usual projection of the elliptic fibrations\footnote{To derive this sequence the flatness of $\pi$ and stability of $V$ are necessary.}. Therefore wherever the rank of the kernel drops, must be the position of the spectral cover. 

This approach, though explicit and computationally tractable, has some drawbacks. The obvious one is that it cannot immediatlely provide information about the spectral sheaf. The other problem is that it is possible and quite common that the spectral cover may wrap components of some non-generic elliptic fibers (i.e. when the restriction of the vector bundle on those non-generic fibers is unstable). In such cases it is possible that the number of global sections of the twisted vector bundle on these fibers jump instead of dropping, and since the algorithm sketched above is designed to detect where the kernel drops, it cannot find these vertical components of the spectral cover\footnote{As long as one wants to find the spectral cover only, it is still possible to use this algorithm, but with other twists to find the missing components. We have employed this technique in recent work \cite{Anderson:2019axt}, but in practice it can be very slow for Calabi-Yau threefolds. }.

To solve the first problem in \cite{Donagi:2011dv}, it was conjectured that the cokernel, $\mathcal{L}$, of the following evaluation map can provide a defining relation for the spectral sheaf,
\begin{eqnarray}
\begin{tikzcd}
0 \arrow[r] & \pi^*\pi_* (V\otimes \mathcal{O}_X(\sigma))  \arrow[r,"ev"] & V\otimes \mathcal{O}_X(\sigma) \arrow[r] & \mathcal{L} \arrow[r] & 0.
\end{tikzcd}
\end{eqnarray} 
However, although $\mathcal{L}$ is supported over the spectral cover, it is not the spectral sheaf generally (in particular when some of the line bundles in the monad have zero relative degree zero).

In our approach, we simply use the resolution of the Poincare sheaf to compute the Fourier-Mukai transforms directly, and is clear from \eref{FMofMonad} that this yields something very similar in spirit to the approaches mentioned above.


\subsection{An extension bundle defined on an elliptic fibration with fibral divisors}
In a similar spirit to the previous sections, it should be noted that a generic bundle chosen over an elliptic threefold with fibral divisors will unfortunately \emph{not necessarily} correspond to a smooth spectral cover with the topology we derived in Section \ref{section3}. However, we can verify that in some simple cases the explicit examples we construct do produce smooth spectral covers with the expected form. Moreover, the techniques outlined in the previous subsections for explicitly computing FM transforms carry over smoothly into this new geometric setting.

For simplicity, we will fix the Calabi-Yau geometry explicitly from the start to be given by an anticanconical hypersurface in the following toric variety:
\begin{eqnarray}
\begin{array}{ccccccc|c}
X & Y & Z & E & x_1 & x_2 & x_3 & p \\
3 & 2 & 1 & 0 & 0 & 0 & 0 & 6 \\
9 & 6 & 0 & 0 & 1 & 1 & 1 & 18\\
8 & 5 & 0 & 1 & 0 & 1 & 1 & 16
\end{array}
\end{eqnarray}
\\
Note that here we denote the single exceptional (i.e. fibral) divisor in this geometry as $E$ and the divisor class of $x_1$ is $D-E$ with $D$ being the hyperplane divisor in the base, $B_2=\mathbb{P}^2$. The image of $E$ on the base is a line homologous to the hyperplane, here denoted $D$. Over $D$ all of the fibers are degenerate of the Kodaira type $I_2$. Also one can show that $E$ satisfies
\begin{eqnarray}
E^2 = -2 \sigma \cdot D + 7 D \cdot E - 6 f.
\end{eqnarray}  
\\

 To illustrate a Fourier-Mukai transform here we can begin by choosing the simple rank two bundle defined by extension of two line bundles chosen in \eref{special_choice} (there in the case of a Weierstrass threefold)
 \begin{eqnarray}
 0 \longrightarrow \mathcal{O}_X(-\sigma+D_b) \longrightarrow V_2 \longrightarrow \mathcal{O}_X(\sigma-D_b) \longrightarrow 0. \nonumber
 \end{eqnarray}
The calculation follows along exactly the same lines as outlined in previous sections, the only interesting point here is the existence of the (-2) curves. As we saw in the Weierstrass case, requiring a non degenerate spectral cover, implies that $2D_b-c_1(B)$ must be effective. So in the present case, the Fourier-Mukai transform of $V_2$ is given by,
\begin{equation}
\Phi(V_2)=\mathcal{O}_{2\sigma+2D_b +c_1(B)}(\sigma+D_b).\nonumber
\end{equation}
In this case, the number of $(-2)$-curves in the spectral cover induced by the exceptional divisor is $\kappa := D\cdot_{B_2} (2D_b+c_1(B))$. So clearly the line bundle over the spectral cover is trivial with respect to the $(-2)$-curves, since $c_1(\mathcal{L})=\sigma+D_b$. 

From this starting point though, it is clear that we choose a new spectral sheaf with some of these exceptional divisors ``turned on", and apply the inverse Fourier-Mukai transform. This will allow us to see how to modify a simple vector bundles line the one above so that its Fourier-Mukai transform will have some non-trivial dependence on the fibral $(-2)$-curves.

To this end, recall that the Fourier-Mukai transform above is given by a short exact sequence, 
\begin{eqnarray}
0\longrightarrow \mathcal{O}_X(-\sigma+K_b-D_b) \longrightarrow \mathcal{O}_X(\sigma+D_b)\longrightarrow \Phi(V_2)\longrightarrow 0. \nonumber
\end{eqnarray}
Now if we twist the above sequence with the $\mathcal{O}_X(E)$, then we obtain a Fourier-Mukai transform of a new stable rank two bundle $\tilde{V}_2$ with spectral line bundle,

\begin{eqnarray}
c_1(\mathcal{L})=\sigma+D_b +\sum_{i=1}^{\kappa} e_i.
\end{eqnarray}
So twisting with $\mathcal{O}_X(E)$ turns on all of the exceptional divisors with multiplicity one. 

Now it is possible to apply an inverse Fourier-Mukai transform. We will omit the details here from brevity and simply state the result, namely a defining sequence for a new bundle $\tilde{V}_2$,
\begin{eqnarray}
0\longrightarrow \mathcal{O}_X(-\sigma+D_b) \longrightarrow \tilde{V}_2 \longrightarrow &\mathcal{O}_X(\sigma-D_b +D-E) \longrightarrow\nonumber\\
&\mathcal{O}_{D-E}(-\sigma+D_b +D+K_B) \longrightarrow 0.
\end{eqnarray}
Note that $D-E$ is an effective divisor. We can easily compute the Chern character of $\tilde{V}_2$ from the exact sequence above (and using GRR),
\begin{eqnarray}
ch(\tilde{V}_2)=2 - \sigma(2 D_b+c_1(B)) +E\cdot (2 D_b+c_1(B))- D_b^2+ D\cdot (K_B-2D_b).
\end{eqnarray}
This is in agreement with the topological equations derived above with $\beta_i =1$, $\kappa=D\cdot(2D_b+c_1(B))$ and $\zeta = - (2D_b+c_1(B))$.

\subsection{A bundle defined via extension on a CY threefold with rk(MW)=1}
Once again in the case of an elliptic manifold with more than one section (and a holomorphic zero-section) we can illustrate the techniques of an FM transform via a simple rank two vector bundle defined via an extension,
\begin{eqnarray}\label{mw_eg}
0\longrightarrow \mathcal{O}_X(-\sigma-S_1+D_b) \longrightarrow V_2 \longrightarrow \mathcal{O}_X(\sigma+S_1-D_b)\longrightarrow 0.
\end{eqnarray}
where here $S_1$ is the Shioda map (see Section \ref{sectionhol}) associated to the second section to the elliptic fibration.

Following the same pattern as in the Weierstrass case, we first compute the extension group,
\begin{eqnarray}
Ext^1(\mathcal{O}_X(\sigma+S_1-D_b),\mathcal{O}_X(-\sigma-S_1+D_b)) = H^1(X,\mathcal{O}_X(-2\sigma-2 S_1+2 D_b)).
\end{eqnarray}
To use Leray spectral sequence we need to know the derived direct images of  $\mathcal{O}_X(-2\sigma_1)$. With the help of Koszul sequence for $\sigma_1$ one obtains
\begin{eqnarray}
R \pi_* \mathcal{O}_X(-2\sigma_1) =( K_B \oplus K_B^{-1} ) [-1].
\end{eqnarray} 

So we see that the extension group decomposes into two subgroups,
\begin{eqnarray}
&Ext^1(\mathcal{O}_X(\sigma+S_1-D_b),\mathcal{O}_X(-\sigma-S_1+D_b)) =\nonumber \\
&H^0(B, \mathcal{O}_B(2D_b+c_1(B))\oplus \mathcal{O}_B(2 D_b + 3 c_1(B))).
\end{eqnarray}

We expect that these two subgroups determine the complex structure of the spectral cover, and if we choose a generic element (assuming $2D_b+3c_1(B)$ is effective), the spectral cover must be smooth, and the topological formulas derived in Section \ref{sectionhol} must be valid. 

Before computing the Fourier-Mukai transform of this bundle, it is useful to consider the Chern character of the bundle given in \eref{mw_eg},
\begin{eqnarray}
ch(V_2)=2-(3 c_1(B)+2 D_b) \sigma - (3 c_1(B)+2D_b) S_1+D_b^2-2 c_1(B)^2.
\end{eqnarray}
From this form, we expect that if the topological formulas given in Section \ref{sectionhol} are satisfied, the divisor class of $S$ must be $2\sigma+2 D_b+3 c_1(B)$, and $c_1(\mathcal{L}) = \sigma-S_1+ c_1(B)+D_b$. 

Now we can compute the Fourier-Mukai explicitly (along the same lines as in previous sections) and obtain
\begin{eqnarray}
\Phi(\mathcal{O}_X(\sigma+S_1-D_b)) &=& \mathcal{O}_X(-\sigma-S_1-2 c_1(B)-D_b), \\
\Phi(\mathcal{O}_X(-\sigma-S_1+D_b)) &=& \mathcal{O}_X(\sigma-S_1+ c_1(B)+D_b)[-1].
\end{eqnarray}
Therefore the Fourier Mukai transform of $V_2$ is simply given by the following torsion sheaf,
\begin{eqnarray}
\Phi(V_2) = \mathcal{O}_{2\sigma + 2 D_b+3 c_1(B)}(-\sigma-S_1+c_1(B)+D_b)[-1].
\end{eqnarray}
In this carefully engineered example then, we are once again able to confirm the results derived in Section \ref{sectionhol}, but we emphasize again that the topological formulas derived will not generally satisfied by a randomly chosen bundle on the elliptic threefold.

\section{Small Instanton Transitions and Spectral Covers}\label{section_small_inst}

An application of the tools we have developed in Sections \ref{section3} is to consider small instanton transitions \cite{Ovrut:2000qi} (i.e. M5-brane/Fixed plane transitions in the language of heterotic M-theory \cite{Horava:1995qa}) involving spectral cover bundles. This subject was first explored in depth in \cite{Ovrut:2000qi,Buchbinder:2002ji} and there a simple form for such transitions were found for smooth spectral covers within Weierstrass models. Within that geometric setting, the authors categorized possible small instanton transitions involving spectral covers as a) Gauge group changing or b) Chirality changing depending on which components of the effective curve class 
\beq
W=W_{B}\sigma + a_f f
\eeq
(wrapped by the 5-brane) are ``absorbed" into the bundle on the a boundary brane. Here $\sigma$ is the holomorphic section of the Weierstrass 3-fold, $W_B$ is a curve within the base $B_2$ and $f$ the fiber class. The authors concluded that in the case that a part of the 5-brane wrapping the fiber class is absorbed into the bundle this can result in case a) above while if a curve in the base is involved (i.e $W_B$ above) then the transition will induce a chirality change in the heterotic effective theory, while in the case of purely ``vertical" transitions (involving detaching a part of $a_f$ above) the chirality is unchanged.

In the following section we will demonstrate that the generalized geometric setting for elliptically fibered CY 3-folds and spectral covers that we have found in Sections \ref{section3} provides new possibilities for such 5-brane transitions. In particular, we will illustrate these possibilities in the case of a transition involving a 5-brane wrapping a curve that is part of a fibral divisor (in the geometric setting of Section \ref{section3})

\subsection{New Chirality Changing Small Instanton Transitions}

Consider for simplicity the case that $X_3$ contains a single fibral divisor class, $D_1$. Suppose that the small instanton is localized on a component of the $I_2$ fibers, $C_1$ (as defined in Section \ref{section3}) with class,
\begin{eqnarray}
[C_1] = (D-D_1)\cdot D
\end{eqnarray}
where $D$ is a divisor pulled back from the base, $B_2$ and $D_1$ is the fibral divisor. Recall that in the case of a CY 3-fold of the type described in Section \ref{section3} we can parameterize the topology of a general bundle $V$ as
\beq\label{top_inst}
ch(\mathcal{E})= N - (\sigma \eta + \omega f + \sum \zeta D_1) + \frac{1}{2} c_3(V)~.
\eeq

As described in \cite{Ovrut:2000qi}, if the 5-brane is moved to touch the $E_8$ fixed plane in a small instanton transition, this geometrically results first in a torsion sheaf $V_{C_1}$ supported over $C_1$, which can be combined with the initial smooth $SU(N)$ bundle $V$ to make a torsion free sheaf $\tilde{V}$:

\begin{eqnarray}\label{Hecke}
\begin{tikzcd}
0\arrow[r]&\tilde{V}\arrow[r] & V\arrow[r] & i_{C_1*} \mathcal{F} \arrow[r]&0,
\end{tikzcd}
\end{eqnarray}
where $i_{C_1}:C_1 \hookrightarrow X$ is the inclusion of the curve mentioned above, and $\mathcal{F}$ is the sheaf supported over the curve $C_1$, wrapped by the 5-brane. The specific order of the sheaves in (\ref{Hecke}) is chosen to describe the \textit{absorption} of the 5-brane. 

The final step in the process of the small instanton transition is to consider, for specific choices for $\mathcal{F}$, whether it is possible to ``smooth out" $\tilde{V}$, to a final smooth/stable vector bundle, $\hat{V}$ as in \cite{Ovrut:2000qi}. To this end, we consider choices of sheaf ${\cal F}$ above (corresponding to parts of the 5-brane class which can be ``detached" and absorbed into $\tilde{V}$) and ask whether the resulting bundle can be smoothed. In the case of the single fibral divisor we are considering (i.e. $I_2$ fibers as in Section \ref{section3}), the curve being wrapped by the 5-branes is topologically a $\mathbb{P}^1$ and we can take the sheaf supported over the 5-brane to be simply a line bundle. Below we explore two choices of this line bundle. \\

\noindent {\bf Case 1: $\mathcal{F} = \mathcal{O}_{C_1}(-1)$}

From \eref{Hecke}, the total Chen character of $\tilde{V}$ is,
\begin{eqnarray}
ch(\tilde{V}) = ch(V) - [C_1] =ch(V) + D\cdot D_1 - f.
\end{eqnarray}
In addition, recall that the Fourier-Mukai transform of $\mathcal{O}_{C_1}(-1)$ is $\mathcal{I}_{C_1} = \mathcal{O}_{C_2}(-2)$. So one can apply the Fourier-Mukai functor to (\ref{Hecke}) to obtain,
\begin{eqnarray}
\begin{tikzcd}
0\arrow[r]& i_{C_2*}\mathcal{O}(-2) \arrow[r] & i_{\tilde{S}} \tilde{\mathcal{L}} \arrow[r] & i_{S} \mathcal{L} \arrow[r] & 0,
\end{tikzcd}
\end{eqnarray}
where $\Phi(V)= i_{S} \mathcal{L}[-1]$ and $\Phi(\tilde{V}) =  \tilde{\mathcal{L}}$ are Fourier-Mukai transforms of $V$ and $\tilde{V}$ which are torsion sheaves supported over the N-sheeted covers of the base, $S$ and $\tilde{S}$ respectively. Taking the case that $S$ is integral, and $C_2$ is one of the $(-2)$-curves which $S$ wraps, then $\tilde{S}=S$, and we get,
\begin{eqnarray}
c_1(\tilde{\mathcal{L}}) = c_1(\mathcal{L}) + e_1.
\end{eqnarray}
Note that $\tilde{\mathcal{L}}$ is singular over $C_2$ ($=e_1$), as may be expected\footnote{Due to the flatness of the projection and the Poincare bundle in the definition of the FM functor we use here, singularity of the ``vector bundle" and the spectral sheaf are closely correlated.}, however, in the process of deforming $\tilde{V}$ to a smooth bundle, $\tilde{\mathcal{L}}$ may also be smoothed out to a line bundle $\hat{\mathcal{L}}$ with the same topology. In this case we can say from the topological data derived earlier in this section that the corresponding (hypothetically) smooth vector bundle $\hat{V}$ must have the following topology (see \eref{top_inst} above)
\begin{eqnarray}
&&\zeta(\hat{V}) = \zeta(V) - D, \\[4pt]
&&\omega(\hat{V}) = \omega(V) + f, \\[4pt]
&&ch(\hat{V}) = ch(V) + D\cdot D_1 - f.
\end{eqnarray}
For these choices, $ch(\hat{V})$ is the same as $ch(\tilde{V})$. So we conclude this transition is topologically unobstructed. In this case we can see that the third Chern character doesn't change in this transition (also $\gamma$ remains unchanged), therefore neither the chiral index or zero-mode spectrum are changed. 
\\

\noindent{\bf Case 2: $\mathcal{F} = \mathcal{O}_{c_1}(-2)$}

As above, from (\ref{Hecke}) we compute the Chern character of $\tilde{V}$ as 

\begin{eqnarray}
ch(\tilde{V}) = ch(V) - [C_1] + 1 w ,
\end{eqnarray}
where $w$ is dual to the zero cycles. Note that if $\tilde{V}$ can be smoothed, we expect $ch(\tilde{V})=ch(\hat{V})$ for the final smooth bundle after the small instanton transition. Thus it is clear that both the second Chern class and chirality can change in this case,
\begin{eqnarray}\label{TopofVhat}
&& c_2(\hat{V})=c_2(V) + D \cdot (D-D_1), \\
&& \frac{1}{2}c_3 (\hat{V}) = \frac{1}{2} c_3(V) +1.
\end{eqnarray}
To address the question of smoothing, we simply apply the Fourier Mukai functor to \eref{Hecke} for the chosen $i_{C_1*} \mathcal{F}$ and assume $V$ is already $WIT_1$,  
\begin{eqnarray}\label{Vhat}
\begin{tikzcd}
0\arrow[r]& i_{\tilde{S}*} \tilde{\mathcal{L}} \arrow[r] & i_{S*} \mathcal{L} \arrow[r]& i_{C_1*}\mathcal{O}_{C_1} \arrow[r]& 0,
\end{tikzcd}
\end{eqnarray}
and we noted that $\Phi(i_{C_1 *} \mathcal{O}_{C_1}(-2)) = i_{C_1*} \mathcal{O}_{C_1}[-1]$. 

Now it must be observed that as long as the above short exact sequence can exist, the sheaf $\tilde{V}$ is indeed $WIT_1$. Note that since an irreducible spectral cover never wraps $C_1$, then the existence of this sequence forces both $S$ and $\tilde{S}$ to have vertical components that contain $C_1$. As a result then, we can choose to consider a small instanton transition in which the spectral cover of the initial bundle $V$ is \emph{reducible}
with vertical (i.e fiber-directions) and horizontal components,
\begin{eqnarray}
S = S_V \cup S_H,
\end{eqnarray}
where $S_V$ contains $C_1$. For simplicity, we will illustrate this transition below in the case that the divisor class $S_V$ is simply $D$, and $\mathcal{L}_V$ is a line bundle. 

Note that although we are choosing the spectral cover to be reducible, it is not the case that $V$ itself must be a reducible bundle. As a next step, we can consider what topological constraints must be in place for a stable degree zero vector bundle such that its Fourier Mukai transform $i_{S*} \mathcal{L}$ is made of a vertical and horizontal piece:
\begin{eqnarray}
\begin{tikzcd}
0\arrow[r]&  i_{S_H*}\mathcal{L}_V \arrow[r]& i_{S*} \mathcal{L} \arrow[r]& i_{S_V*}\mathcal{L}_H\arrow[r] & 0.
\end{tikzcd}
\end{eqnarray} 
Following the same procedure as before we can derive the the topological data,

\begin{eqnarray}
&& [S_H] = N \sigma + \eta - D, \\
&& [S_H]\cdot \left( c_1(\mathcal{L}_H) -\frac{1}{2} [S_H] \right) + D\cdot \left(c_1(\mathcal{L}_V)-\frac{1}{2} D \right) \nonumber \\
&& = (N \sigma +\eta) \left(-\frac{1}{2} c_1(B)\right) - \frac{1}{2} c_3(V) f- \zeta e_1.
\end{eqnarray}
A solution for this equation can be given as,
\begin{eqnarray}
&& c_1(\mathcal{L}_H) = -\frac{1}{2} \left( c_1(B) - [S_H] \right) + \gamma_H , \\
&& \gamma_H = \lambda_H (N \sigma -\eta + D + N c_1(B)) + \delta \sigma, \\
&& c_1(\mathcal{L}_V) = -\zeta e + \lambda_V D - \delta \sigma, \\
&& \frac{1}{2} c_3(V) = \lambda_H \eta (\eta - N c_1(B)) - \lambda_V + \frac{1}{2}D\cdot (D-c_1(B)). 
\end{eqnarray}
After a tedious algebraic calculation, one can derive a formula for $\omega$, but it is not necessary here. Finally if we require both $V$ and $\tilde{V}$ have the same spectral cover\footnote{Note for simplicity we assumed $\mathcal{L}_H$ is independent of the $(-2)$-curves on the horizontal components $S_H$. }, then (\ref{Vhat}) implies the following relation between the vertical parts of the spectral sheaves,
\begin{eqnarray}
\tilde{\mathcal{L}}_{\tilde{V}} =\mathcal{L}_{V} \otimes \mathcal{O}_{S_V}(-D+E).
\end{eqnarray}
Therefore we easily get the following relations between the parameters of $\tilde{V}$ and $V$,
\begin{eqnarray}
&& \lambda_{\tilde{V}} = \lambda_V -1, \\
&& \zeta_{\hat{V}} = \zeta_{V} -1.
\end{eqnarray}
Moreover if we put $\delta_V = \delta_{\hat{V}}=0$, we can see by the above arguments that,
\begin{eqnarray}
\omega_{\hat{V}} = \omega_V + 1
\end{eqnarray}

Finally, we arrive at a point where we can compare the above conditions on $V$ and $\tilde{V}$ with the relations \eref{TopofVhat} derived before and observe that they are exactly the same. Thus, the transition is unobstructed and we have provided an example of a complete (i.e. smooth-able) chirality changing transition involving fibral curves. 

We should emphasize that the above geometry is by no means general and many choices were made for simplicity of computation. None-the-less, it serves to illustrate that the existence of fibral divisors in the elliptically fibered CY 3-fold will make new forms of small instantons possible. In particular, the example above is a chirality changing transition that is unique compared to those classified in \cite{Ovrut:2000qi} for Weierstrass form (in which ``vertical" transitions changed only the gauge group and ``horizontal" curves led to chirality change). In this example we find chirality change from new vertical curves for the 5-brane to wrap and the gauge group remains unchanged even though $C_1$ is a vertical curve.  

\section{Reducible spectral covers and obstructions to smoothing}\label{section_obstruction}

As illustrated by the examples in Section \ref{section_egs}, there are many limitations to the analysis that we completed in Sections \ref{section2} to \ref{sectionhol}.  First, the Picard number of the spectral cover maybe larger $1+h^{1,1}(B_2)$ generically. This corresponds to spectral surfaces in which there exist more divisors than those inherited from the ambient Calabi-Yau threefold. Moreover, it is known that at higher co-dimensional loci in moduli space, this Picard group can in fact jump \cite{Amerik1998}. Second, as seen in the examples in previous sections, the spectral cover can be singular, and therefore one cannot predict the general form of $ch(i_{S*}\mathcal{L})$. 

In these cases it may be possible to choose special sheaves $\mathcal{L}$ that ``obstruct" the deformation of the spectral cover to a smooth one. In other words, the corresponding vector bundles lands on a different component\footnote{Note that this cannot happen for a vector bundle over an elliptically fibered K3 surface. This phenomenon only appears for CY manifolds of complex dimension 3 or higher.} than the one that is analyzed in \cite{Friedman:1997yq,Friedman:1997ih}. In this section we briefly outline how such a situation might be realized in the case that spectral cover is reducible but reduced. This analysis has some similarity to examples analyzed in \cite{Donagi:2011jy}.   

We begin with the spectral data $(\mathcal{L},S)$ of a bundle $V$ defined over a Weierstrass CY threefold $\pi : X \rightarrow B$, where
\begin{eqnarray}
&&S := S_1 \cup_{\Sigma} S_2, \\
&&0\longrightarrow \mathcal{L}_1 \longrightarrow \mathcal{L} \longrightarrow \mathcal{L}_2\longrightarrow 0.
\end{eqnarray}
As usual 
\begin{eqnarray}
&&ch(V) = N-( \sigma \eta + \omega) + \frac{1}{2} c_3(V), \\
&&- ch(\Phi(V)) = ch(\mathcal{L}) = (N\sigma+\eta) + (N \sigma+\eta) (-\frac{c_1(B)}{2})  + \left( \frac{1}{6} n c_1(B)^2 - \omega \right).
\end{eqnarray}

Now we assume that  
\begin{eqnarray}
&&[S_1] = n_1 \sigma + \eta_1 , \\
&&[S_2] = n_2 \sigma + \eta_2 , \\
&&N = N_1+N_2 , \\
&&\eta =\eta_1 +\eta_2 .
\end{eqnarray}

With these assumptions, the general for for $c_1(\mathcal{L}_1)$ and $c_1(\mathcal{L}_2)$ are given below,
\begin{eqnarray}
c_1(\mathcal{L}_1) &=& \frac{1}{2} (-c_1(B) +[S_1]) + \gamma_1 + \alpha_1 [S_2] , \\
c_1(\mathcal{L}_2) &=& \frac{1}{2} (-c_1(B) +[S_2]) + \gamma_2 + \alpha_2 [S_1] , \\
\gamma_i &=& \lambda_i  (N_i \sigma-\eta_i+N_i c_1(B)) , \\
\frac{1}{2} c_3(V) &=& \sum_i N_i \lambda_i \eta_i (\eta_i-N_i c_1(B)).  
\end{eqnarray}

The main difference of the equations above with the standard one is the existence of the terms $ \alpha_1 [S_2]$  and $ \alpha_2 [S_1]$. For consistency we demand,
\begin{eqnarray}
\alpha_1 +\alpha_2 = 0.
\end{eqnarray}
Note the existence of such terms implies $(\mathcal{L}_i,S_i)$ are spectral data of vector bundles $V_i$ with first Chern class,
\begin{eqnarray}
c_1(V_i)= \alpha_i (N_1 \eta_2+N_2 \eta_1-N_1 N_2 c_1(B)).
\end{eqnarray}
It is next possible to compute $\omega$ as before,
\begin{eqnarray}
&&\frac{1}{6} N c_1(B)^2 - \omega = \nonumber \\
&&\frac{N c_1(B)^2}{8}+\frac{c_1(B)^2}{24}(N_1^3+N_2^3)+\frac{1}{8} \left( N_1 \eta_1 (\eta_1-N_1 c_1(B)) + N_2 \eta_2 (\eta_2-N_2 c_1(B)) \right)  \nonumber \\
&&+\frac{1}{2} \pi_{1*} \gamma_1^2 +\frac{1}{2} \pi_{2*}\gamma_2^2 \nonumber \\
&&+\frac{1}{2} \Sigma \cdot \left( \alpha_1^2 [S_2]+\alpha_2^2 [S_1]+2 \alpha_1 \gamma_1 +2 \alpha_2 \gamma_2 \right).
\end{eqnarray}

After some algebra it can be shown that \emph{only for $\alpha_1=-\alpha_2 = \pm \frac{1}{2}$} can the above equation be simplified to,
\begin{eqnarray}
&&\frac{1}{6} N c_1(B)^2 - \omega = \nonumber \\
&&\frac{N c_1(B)^2}{8}+\frac{c_1(B)^2}{24}N^3+\frac{1}{8}  N \eta (\eta-N c_1(B))  \nonumber \\
&&+\frac{1}{2} \pi_{1*} \gamma_1^2 +\frac{1}{2} \pi_{2*}\gamma_2^2 \nonumber \\
&&+\Sigma \cdot \left( \alpha_1 \gamma_1 + \alpha_2 \gamma_2 \right).
\end{eqnarray}
This is almost the same as the standard formula expected from Section \ref{section2} if there exists a $\lambda$ such that 
\begin{eqnarray}\label{top}
&&\frac{1}{2}\pi_* \gamma^2 = \nonumber \\
&&\frac{1}{2} \pi_{1*} \gamma_1^2 +\frac{1}{2} \pi_{2*}\gamma_2^2 + \nonumber \\
&&\Sigma \cdot \left( \alpha_1 \gamma_1 + \alpha_2 \gamma_2 \right).
\end{eqnarray}

We come now to our central claim in this section: \\

\noindent \emph{If restriction of $\mathcal{L}$ on $\Sigma$ is a trivial line bundle, then it is always possible to deform the ``singular" spectral data to a ``smooth" spectral data, such that it satisfies the generic formulae expected in \eref{fmw1} -- \eref{chern_spec}. Otherwise it is impossible (generically). In particular if the restriction is a non-trivial degree zero line bundle, the deformation is obstructed.}\\

\vspace{3pt}

First note that if $\mathcal{L}$ is defined as 
\begin{eqnarray}
0\longrightarrow \mathcal{L}_1 \longrightarrow \mathcal{L} \longrightarrow \mathcal{L}_2\longrightarrow 0,
\end{eqnarray}
the restriction of $\mathcal{L}$ on $S_1$ and $S_2$ are 
\begin{eqnarray}
&&\mathcal{L}_1 \otimes K_{S_2}|_{S_1}, \nonumber \\
&&\mathcal{L}_2,
\end{eqnarray} 
respectively. Therefore the line bundle induced over $\Sigma$ lives in
\begin{eqnarray}\label{ext1}
Hom_{\Sigma} (\mathcal{L}_2,\mathcal{L}_1 \otimes K_{S_2}|_{S_1}) \simeq Ext^1_{X} (i_{S_2*} \mathcal{L}_2,i_{S_1*} \mathcal{L}_1),
\end{eqnarray}
corresponding to extensions. Conversely, if we define $\mathcal{L}$ as,

\begin{eqnarray}
0\longrightarrow \mathcal{L}_2 \longrightarrow \mathcal{L} \longrightarrow \mathcal{L}_1\longrightarrow 0,
\end{eqnarray}

the restriction of $\mathcal{L}$ on $S_1$ and $S_2$ are 
\begin{eqnarray}
&&\mathcal{L}_2 \otimes K_{S_1}|_{S_2}, \nonumber \\
&&\mathcal{L}_1,
\end{eqnarray} 
respectively. Therefore the line bundle induced over $\Sigma$ lives in
\begin{eqnarray}\label{ext2}
Hom_{\Sigma} (\mathcal{L}_1,\mathcal{L}_2 \otimes K_{S_1}|_{S_2}) \simeq Ext^1_{X} (i_{S_1*} \mathcal{L}_1,i_{S_2*} \mathcal{L}_2),
\end{eqnarray}
corresponding to the opposite extensions. If we rewrite the left hand side of \eref{ext1} as,
\begin{eqnarray}
&&H^0(\Sigma,\mathcal{F}), \nonumber \\
&&\mathcal{F} := \mathcal{L}_1\otimes \mathcal{L}^*_2 \otimes K_{S_2}|_{S_1},
\end{eqnarray}
then \ref{ext2} can be written as, 
\begin{eqnarray}
H^0(\Sigma,\mathcal{F}^* \otimes K_{\Sigma}).
\end{eqnarray}
Therefore we see\footnote{We could also choose $\mathcal{F}^*\otimes K_{\Sigma} \simeq \mathcal{O}_{\Sigma}$. But since the analysis would run along very similar lines, we choose to just focus on the first case.} if $\mathcal{F}\simeq \mathcal{O}_{\Sigma}$ ,then both extensions are possible, and we can deform the spectral data to generic ``smooth" one described in FMW. 

We can indeed check that in this case there is a $\lambda$ that satisfy \eref{top}. To show that we choose,
\begin{eqnarray}
\alpha_1=-\alpha_2 =-\frac{1}{2}
\end{eqnarray}
(the other choice corresponds to $\mathcal{F}\otimes K_{\Sigma} \simeq \mathcal{O}_{\Sigma}$). Notice in this case if $\gamma_1 = \gamma_2$ as a divisor in $X$ then $\mathcal{F}\simeq \mathcal{O}_{\Sigma}$. This constraint is equivalent to,

\begin{eqnarray}
N_1 \lambda_1 =N_2 \lambda_2 , \\
\eta_1 \lambda_1 =\eta_2 \lambda_2.
\end{eqnarray}
Let us look at \ref{top} more closely,
\begin{eqnarray}
&& \frac{1}{2}\lambda^2 N\eta (\eta- N c_1(B))  = \nonumber \\
&& \frac{1}{2}\lambda_1^2 N_1 \eta_1 (\eta_1- N_1 c_1(B)) - \frac{1}{2}\lambda_1 N_2 \eta_1 (\eta_1- N_1 c_1(B)) +\nonumber\\
&& \frac{1}{2}\lambda_2^2 N_2 \eta_2 (\eta_2- N_2 c_1(B)) +\frac{1}{2}\lambda_2 N_1 \eta_2 (\eta_2- N_2 c_1(B)).
\end{eqnarray}
The second terms in the 2nd and 3rd line cancel. To find $\lambda$ we choose an ansatz $\lambda=\alpha \lambda_1 \lambda_2$, and use the constraints above, we can see the solution is ,
\begin{eqnarray}
\lambda=\frac{\lambda_1 \lambda_2}{\lambda_1+\lambda_2}.
\end{eqnarray}
On the other hand if we request $\gamma_1 = \gamma_2$ only over $\Sigma$, i.e.

\begin{eqnarray}
S_1 \cdot S_2 \cdot \gamma_1 = \gamma_1|_{\Sigma} = \gamma_2|_{\Sigma}  =S_1 \cdot S_2 \cdot \gamma_2,
\end{eqnarray}
then it means  $\mathcal{F}$ is an element of $J(\Sigma)$ but it is not necessarily a trivial line bundle (as $g(\Sigma)\geq1$ generally). In this case there is no solution for $\Lambda$ generally. 

In summary then, we have seen in this section that the properties of reducible spectral covers may indeed be quite distinct from their smooth cousins. 

\section{Conclusions and future directions}\label{section_conclusions}
In this work we have generalized the famous spectral cover construction of Friedman, Morgan and Witten \cite{Friedman:1997yq,Friedman:1997ih,Curio:1998vu} to the case of elliptic Calabi-Yau threefolds with higher rank Picard group (i.e. containing either fibral divisors or multiple sections to the elliptic fibration). In particular, the well-established work of \cite{Friedman:1997yq,Curio:1998vu} provided a simple formula for the Chern classes of bundles associated to smooth (i.e. reduced and irreducible) spectral covers in Weierstrass CY 3-folds:
\begin{align}
& c_1(\mathcal{E})=0 \\
& c_2(\mathcal{E})=\eta \sigma - \frac{N^3-N}{24}c_1(B_2)^2+\frac{N}{2}\left(\lambda^2-\frac{1}{4}\right)\eta \cdot \left(\eta-Nc_1(B_2) \right) \\
& c_3(\mathcal{E})=2\lambda \sigma \eta \cdot \left(\eta-Nc_1(B_2) \right)
\label{chern_spec2}
\end{align}
In this work we have utilized the techniques of Fourier-Mukai functors to generalize these formula to bundles defined over geometries with fibral divisors and higher rank Mordell-Weil. In the case of $I_n$ type singular fibers we find that $c_1(\mathcal{E})$ and $c_3(\mathcal{E})$ are unchanged and in the case of $I_2$ fibers we find a correction to the second Chern class of the form
\beq
c_2(\mathcal{E})= \sigma \cdot \eta + \omega_{std}+ (\zeta_1 \cdot \mathcal{S} + \sum_{i=2}^k \beta_{i} )^2 + \sum_{i=2}^k \beta_i^2 + \zeta_1 \cdot D_1
\eeq
where $D_1$ is the new fibral divisor, $\zeta_1$ is an effective class pulled back from the base, $B_2$, $\beta_i$ are integers and the divisor $\mathcal{S}$ is a component of the discriminant locus of the fibration (supporting the $I_2$ fibers) in the base. Here
\beq 
\omega_{std}=- \frac{N^3-N}{24}c_1(B_2)^2+\frac{N}{2}\left(\lambda^2-\frac{1}{4}\right)\eta \cdot \left(\eta-Nc_1(B_2) \right) 
\eeq
Similarly, in the case of an additional, holomorphic zero section we find
\beq
c_2(\mathcal{E})= \sigma \cdot \eta - \beta_1 (\eta+N D_{11}) \cdot S_1 + \left(\omega_{std} -\frac{1}{2} \beta_1^2 (\eta+N D_{11})S_1^2\right) f
\eeq
where $\beta_1$ is integer, $S_1$ is the Shioda map of the new section and $D_{11}$ is a divisor in $B_2$ determined by the triple intersection numbers involving the sections.

In the case that the additional sections are rational rather than holomorphic (and hence can wrap reducible components of fibers over higher-codimensional loci in the base), there remain open questions about how best to define a Fourier-Mukai functor that can accommodate the singular fibers (and a section which wraps some of them). As a result, we cannot yet determine how these topological formula will change. However, we are able to see in this case that interesting new results are possible since we expect not only the second Chern class, but the chiral index to change as well. We have outlined in this work several ways forward on this important problem and we hope to return to it in future work. 

Within heterotic/F-theory duality, the constrained geometric arena --i..e Weierstrass from for both the heterotic and F-theory Calabi-Yau backgrounds -- has long been a frustrating obstacle to studying new phenomena. Within heterotic effective theories for example, there are a number of interesting effects that are believed to have interesting F-theory duals, including perhaps novel mechanisms for moduli stabilization such as the linking of bundle and complex structure moduli in the heterotic theory through the condition of holomorphy\cite{Anderson:2014xha,Anderson:2013qca,Anderson:2011ty,Anderson:2011cza} and potentially new 4-dimensional ${\mathcal N}=1$ dualities including heterotic threefolds admitting multiple elliptic fibrations (and hence leading to multiple, related dual F-theory fourfolds) \cite{Anderson:2016cdu,Anderson:2016ler,Anderson:2010mh}, the F-theory duals of heterotic target space duality \cite{Anderson:2019axt} or F-theory duals \cite{Braun:2018ovc,Clemens:2019dts} of known ``standard model like" heterotic compactifications (including \cite{Anderson:2013xka}). However in all cases, these theories have crucially involved decidedly non-Weierstrass geometry on the heterotic side. These questions have formed the motivation for the present work. We believe that here we have taken important first steps towards extending the geometries for which explicit heterotic/F-theory duals can be constructed.

There remain however, important open questions. First, as mentioned above, we require new and more robust tools to address the general case of a higher rank Mordell-Weil group with rational generators studied in Section \ref{rationalsec}. In addition, as illustrated in the explicit examples constructed in Section \ref{section_egs} all the formulas we have derived in this work have been limited by the restriction of \emph{smoothness} of the spectral cover. In general many examples in the literature (see e.g. \cite{Aspinwall:1998he}) have demonstrated that smooth vector bundles do not necessarily correspond to smooth spectral covers. Indeed, this observation has been a powerful tool in determining the effective physics of T-brane solutions in F-theory \cite{Anderson:2013rka,Cecotti:2010bp,Anderson:2017rpr,Anderson:2017zfm}. By placing the constraint of smoothness on the spectral data, we are clearly loosing information about general components of the bundle moduli space (as illustrated in Section \ref{section_obstruction}). Finally, there remain interesting open questions about how to determine the full Picard groups of spectral covers (since these are surfaces of general type, this is a notoriously hard problem in algebraic geometry, see e.g. \cite{Anderson:2016kuf}) and a number of interesting possibilities remaining to be explored related to higher co-dimensional behavior in moduli spaces (i.e. so-called ``jumping" phenomena or Noether-Lefschetz problems \cite{Donagi:2009ra}). 

One approach to the problem of singular covers above might arise through a recursive approach.
As noted above, the only general topological formulas derived (here and in the literature overall) are for vector bundles realized (modulo the Picard number problem) by smooth spectral covers. In the case that the spectral cover is a union of several components which can be smooth, or non reduced or vertical the main obstacle is providing a general form for the Chern character of the spectral sheaf (which is clearly a hard problem in the algebraic geometry of singular surfaces). However, we might hope to avoid this difficult question by deriving a ``recursive" algorithm to resolve the singularities of the spectral cover that could work in general. For example, if the spectral cover is degenerate, it is still possible to find a locally free resolution (with length one) of the spectral sheaf. We might hope to use Fourier-Mukai transforms to study the vector bundles associated to this resolution. If one can argue that the ``degree of the degeneracy" drops in each step, then this process will terminate at some point.  

All of these problems deserve further attention and are necessary for a general study fo heterotic/F-theory duality. We hope to continue to explore them in future work.

\section*{Acknowledgements}
The authors would like to thank A. Caldararu, P. Oehlmann, A.C. Lopez-Martin, and D.H. Ruiperez for useful discussions. In addition, LA and MK gratefully acknowledge the hospitality of the Simons Center for Geometry and Physics (and the semester long program, ``The Geometry and Physics of Hitchin Systems") during the completion of this work. The work of LA is supported by NSF grant PHY-1720321.

\appendix

\appendix
\section{Basics about derived category}\label{Appendix A}

Since the Fourier-Mukai functor, which we use a lot in this paper, is a special integral transform, we devote this appendix on reviewing some key points about them. For more details, look at \cite{BBRH,Huybrechts}. 
\\

\begin{labeling}{hom}
\item [\textbf{$\mathbf{Hom}_{\mathcal{A}}$}]
 First of all note that any functor between two categories $F: \mathcal{A}\rightarrow \mathcal{B}$ induces a map between the space of morphisms, 

\begin{equation}\label{hommap}
Hom_{\mathcal{A}}(A,B) \rightarrow Hom_{\mathcal{B}} (F(A),F(B)),
\end{equation} 
 
where $A$, $B$ are arbitrary objects of the category $\mathcal{A}$ (i.e. the map is "functorial"). In case the categories are additive the set of morphisms form an abelian group, and in the cases we are concerned in this paper they are actually $\mathbb{C}-$vector spaces. Abelian categories are particular additive categories that for any functor one can define kernel and cokernel. The specific category we need in this paper is $\mathbf{Coh}(X)$, i.e. the category of coherent sheaves over a variety $X$, and the categories derived from that. 
\item [\textbf{Fully faithful functor}]
A functor $F:\mathcal{A}\rightarrow\mathcal{B}$ is called full if the map (\eref{hommap}) is surjective and it is called faithful if it is injective. So a fully faithful functor induces an isomorphism in (\eref{hommap}). 

\item [\textbf{Left and right adjoint}]
A functor $G: \mathcal{B}\rightarrow \mathcal{A}$ is a right adjoint of $F:\mathcal{A}\rightarrow \mathcal{B}$, written as $F \dashv G$ if 
\begin{equation}
Hom_{\mathcal{B}}(F(A),B) \sim Hom_{\mathcal{A}}(A,G(B)),
\end{equation}

where $A\in \mathcal{A}$ and $B\in \mathcal{B}$ are any arbitrary object. In particular one can see 

\begin{equation}
Hom_{\mathcal{B}}(F(A),F(B)) \sim Hom_{\mathcal{A}}(A,GoF(B)), \nonumber
\end{equation} 

\item [\textbf{Equivalence of categories}]

A functor $F: \mathcal{A}\rightarrow \mathcal{B}$ is called equivalence if there are functors $G,H : \mathcal{B}\rightarrow \mathcal{A}$ such that they satisfy the functor isomorphisms $GoF \sim id_{\mathcal{A}}$ and $FoH \sim id_{\mathcal{B}}$. \\
It is now easy to see \cite{Huybrechts} that if a functor is fully faithful and have both left and right adjoint then it is an equivalence.

\item [\textbf{Category of complexes}] Suppose $\mathcal{A}$ is an abelian category. Then one defines the category of complex $C(\mathcal{A})$, which it's objects are complexes of objects in $\\mathcal{A}$,

\begin{equation}
A^{\bullet} := \dots \longrightarrow A^{i-1} \stackrel {d^{i-1}}{\longrightarrow} A^i \stackrel{d^{i}}{\longrightarrow} A^{i+1} \longrightarrow \dots 
\end{equation}
such that $d^i \circ d^{i-1}=0$. The morphisms in $C(\mathcal{A})$ between two objects $h: A^{\bullet} \rightarrow B^{\bullet}$  are defined by a collection of morphisms $\lbrace h^i \rbrace$ in $\mathcal{A}$ as,

\begin{equation}\label{complex morphism}
\begin{tikzcd}
\dots \arrow[r] 
& A^{i-1}\arrow[d,"h^{i-1}"] \arrow[r, "d_A^{i-1}"] 
& A^{i} \arrow[d,"h^{i}"] \arrow[r] 
& \dots \\
\dots \arrow[r] 
& B^{i-1} \arrow[r, "d_B^{i-1}"] 
& B^{i} \arrow[r] 
& \dots
\end{tikzcd}
\end{equation}
which must be commutative. There are several remarks that must be mentioned,\\
 i) One can define the shift functor ,$T: C(\mathcal{A})\rightarrow C(\mathcal{A})$, naturally in this category as,
 
 \begin{eqnarray}
& A^{\bullet}[1]:=T(A^{\bullet}), \nonumber\\
& ( A^{\bullet}[1])^i=A^{i+1}, \qquad d_{A^{\bullet}[1]}^{i}= - d_{A^{\bullet}}^{i+1}.
 \end{eqnarray}\\
 ii) As usual one can define cohomology for complexes,
 \begin{equation}
 \mathcal{H}^i (A^{\bullet})=\frac{Ker (d^{i})}{Im (d^{i-1})}.
 \end{equation}
 Two complexes $A^{\bullet}$, $B^{\bullet}$ are said to be\textbf{ \emph{Quasi Isomorphic}} if all of their cohomologies are isomorphic.\\
 
\item [\textbf{Derived category}] Roughly speaking, derived category is ``derived" from the homotopy category \footnote{Homotopy Category is derived from category of complexes by taking quotient relative to the homotopy equivalence relation.} by localizing with the "ideal of quasi isomorphisms". In other words $Ob(D(\mathcal{A})) := Ob(C(\mathcal{A}))$, and morphisms in $D(\mathcal{A})$ between two objects $A^{\bullet}$, $B^{\bullet}$ are like,
\begin{equation}\label{derived morphism}
\begin{tikzcd}
 & C^{\bullet} \arrow[dl, "qis" '] \arrow[dr, "f"] & \\
 A^{\bullet} & & B^{\bullet}
\end{tikzcd}
\end{equation} 

In general $f$ is a general morphism in homotopy category. As a result if $f$ is also a quasi isomorphism, then the corresponding morphism in the derived category is isomorphism. So \emph{ in $\mathcal{A}$, if cohomology of two complex is isomorphic, then the complexes themselves are isomorphic}. 

\textbf{Note:} From now on we restrict ourselves to bounded derived categories, $D^b(\mathcal{A})$, which it's objects are isomorphic to complexes with bounded cohomology complexes. 
\\

\item [\textbf{Derived functor}] If a functor $F: K(\mathcal{A})\rightarrow K(\mathcal{B})$ between homotopy categories is compatible with quasi isomorphisms, i.e. it sends quasi isomorphisms to quasi isomorphisms (pr equivalently it sends acyclic complexes to acyclic complexes), then it naturally induces a functor on derived categories. But generally it may not happen, so one need to `derive' a functor from $F$ such that it is compatible with `localization' of morphisms with quasi isomorphisms. This functor is called derived functor $RF$. Here we briefly describe the derived functors that we are going to use them in this paper. For general discussions the reader can consult with \cite{BBRH}.   

From now on, we restrict ourselves with categories of coherent sheaves $Coh(X)$ and quasi coherent sheaves $Qcoh(X)$ over a variety $X$. In particular it is possible to show \cite{BBRH} 
\begin{eqnarray}
D^b_{Coh(X)}(Qcoh(X)) \sim D^b(Coh(X)), 
\end{eqnarray}
where the left hand side corresponds to derived category of complexes of quasi coherent sheaves which their cohomologies are coherent sheaves. One define the bounded derived category of $X$ as $D^b(X) :=  D^b(Coh(X))$.

\item [\textbf{Derived direct image}] Here the goal is to find the derived functor of $f_*: Coh(X) \longrightarrow Coh(Y)$ induced from a projective (or at least proper) morphism of varieties $f: X\longrightarrow Y$. 

 If we have proper morphism of varieties $f: X\longrightarrow Y$, then the (right) direct image $Rf_* :D^b(X)\longrightarrow D^b(Y)$ is defoned in the following way,

\indent 1) For any complex of coherent sheaves $A^{\bullet}$ with bounded cohomology, we have an injective resolution $A^{\bullet} \longrightarrow I(A^{\bullet})$. \\
\indent 2) Define
\begin{eqnarray}
Rf_* (A^{\bullet}) :=f_* (I(A^{\bullet})), \nonumber\\
R^if_* (A^{\bullet}) := \mathcal{H}^i(f_* (I(A^{\bullet}))).
\end{eqnarray}  

\item [\textbf{Derived $Hom$ functor and $Ext$ groups}]  Lets start by the following definition,

  \begin{definition} 
  A complex in $\mathcal{I}^{\bullet} \in C(\mathcal{M}od(X))$ is called injective \textsc{complex} if the right exact functor $Hom^{\bullet}_{ C(\mathcal{M}od(X))}(\dots,\mathcal{I}^{\bullet} ): C(\mathcal{M}od(X)) \longrightarrow \mathcal{A}b$ maps any acyclic complex to another acyclic complex (or equivalently map any quasi isomorphism to another quasi isomorphism). \\
  \end{definition}

 Now it can be proved a bounded bellow complex of injective sheaves is actually an injective complex. So as before for a complex $A^{\bullet}$ one can define a resolution by injective objects $B^{\bullet} \rightarrow \mathcal{I}^{\bullet}$, and define 

\begin{eqnarray}
RHom^{\bullet}_{C(\mathcal{M}od(X))}(A^{\bullet},\dots) : D^b(X) \longrightarrow D^b(\mathcal{A}b),\\
RHom^{i}_{C(\mathcal{M}od(X))}(A^{\bullet},B^{\bullet}) : =\mathcal{H}^i (Hom_{C(\mathcal{M}od(X))}(A^{\bullet},\mathcal{I}^{\bullet})).
\end{eqnarray}
Without getting into more details, we state that relative to the first ``variable" (i.e. $A^{\bullet}$), the functor defined above is consistent with the quasi isomorphisms. So if we consider $R Hom$ as a functor on the first variable, it naturally induces a well defied functor in the derived category. Therefore,
\begin{equation}
R Hom : D^0(X) \times D^b(X) \longrightarrow D(\mathcal{A}b),
\end{equation}
where $D^0(X)$ is the opposite category of $D(X)$. 
\begin{definition}
$Ext^i_{D(X)}(A^{\bullet},B^{\bullet}) := R^i Hom (A^{\bullet},B^{\bullet})$.
\end{definition}
 
 So far we only considered the global $Hom$ functor, but in the case of sheaves one can define a local version
\cite{Hartshorne}$\mathcal{H}om$,
 
 \begin{equation}
 R \mathcal{H}om_{\mathcal{O}_X} :  D^0(X) \times D^b(X) \longrightarrow D^b(X),
 \end{equation}
and similar to the global version one has local ``ext" sheaves,

\begin{equation}
\mathcal{E}xt^i_{\mathcal{O}_X}(A^{\bullet},B^{\bullet}) := R^i \mathcal{H}om_{\mathcal{O}_X}(A^{\bullet},B^{\bullet}).
\end{equation}

\item [\textbf{Derived tensor product}] Lets start by reviewing some standard facts,\\

\indent i) For any sheaf $A$, the functor $A\otimes \dots$ is right exact, and $A$ is flat if $A\otimes \dots$ is exact.\\
\indent ii) For any coherent sheaf $A$, there is a flat resolution of finite length 
\begin{equation}
\dots \longrightarrow \mathcal{F}_1\longrightarrow \mathcal{F}_0 \longrightarrow A \longrightarrow 0, 
\end{equation}
where $\mathcal{F}_i$'s are flat sheaves. \\
\indent iii) One can define the tensor product of two complexes $A^{\bullet}\otimes B^{\bullet}$ as a double complex.\\ 
\indent iv) A flat complex is defined as complex $\mathcal{P}^{\bullet}$, which the functor $\mathcal{P}^{\bullet} \otimes \dots$, maps acyclic complexes to acyclic complexes (or equivalently quasi isomorphism to quasi isomorphism).\\
\indent v) A bounded above (in particular bounded) complex of flat sheaves is a flat complex. For a bounded complex of coherent sheaves, $B^{\bullet}$, then (using point (ii) ) one can find a quasi isomorphism $\mathcal{P}^{\bullet} \longrightarrow B^{\bullet}$.   If $\mathcal{P}^{\bullet}$ is both flat and acyclic, then $B^{\bullet} \otimes \mathcal{P}^{\bullet}$ is again acyclic for any complex $B^{\bullet}$.\\

As before one can define the derived tensor product as,

\begin{eqnarray}
RF_{A^{\bullet}} := A^{\bullet} \otimes^{L} \dots : D^b(X) \longrightarrow D^{b}(X), \\
RF^i_{A^{\bullet}}(B^{\bullet}) = \mathcal{H}^i (A^{\bullet} \otimes \mathcal{P}^{\bullet}).
\end{eqnarray} 

Note that the process of defining derived tensor product is symmetric, and one could define it using the first variable. Also if there is a quasi isomorphism $A^{\bullet} \stackrel{qis} \longrightarrow B^{\bullet} $, then we have a functor isomorphism $F_{A^{\bullet}} \sim F_{B^{\bullet}}$. So naturally the derived tensor product descends to a well defined functor in derived category relative to the first variable,
\begin{equation}
\dots \otimes^L \dots : D^b(X) \times D^b(X) \longrightarrow D^b(X).
\end{equation}

\begin{definition}
\begin{equation}
\mathcal{T}or_{i}  (A^{\bullet}, B^{\bullet}):= \mathcal{H}^{-i} (A^{\bullet} \otimes^L B^{\bullet}).
\end{equation}
\end{definition}

\item [\textbf{Derived pullback}] Finally we are at the position to review the definition the left derived functor for the pullback of a morphism $f: (X,\mathcal{O}_X) \longrightarrow (Y,\mathcal{O}_Y)$.  As before, we recall some basic facts and then compare with the general definition. \\
\indent i) Recall that the pull back of a sheaf under $f$ is defined as,
\begin{equation}
f^* (\mathcal{F}) := \mathcal{O}_X \otimes_{f^{-1}\mathcal{O}_Y} f^{-1} \mathcal{F}. 
\end{equation} 
\indent ii) There is a projective resolution for every coherent sheaf,
\begin{equation}
\dots \longrightarrow \mathcal{P}^1 \longrightarrow \mathcal{P}^0 \longrightarrow \mathcal{F} \longrightarrow 0,
\end{equation}

This induces a quasi isomorphism for any bounded complex of coherent sheaves (at least bounded above) one gets a quasi isomorphism $\mathcal{P}^{\bullet}\stackrel{qis}\longrightarrow\mathcal{F}^{\bullet}$.  \\

So by combining these facts and what we learned for derived tensor product we can write,

\begin{eqnarray}
Lf^{*} (\mathcal{F}^{\bullet}) := \mathcal{O}_X \otimes_{f^{-1}\mathcal{O}_Y}^{L} f^{-1} \mathcal{F}^{\bullet},\nonumber\\
L_i f^{*}(\mathcal{F}^{\bullet}) :=\mathcal{H}^{-i} (f^{*} (\mathcal{P}^{\bullet})).
\end{eqnarray}

\item [\textbf{Important identities}] Here we collect the identities that are going to be useful in the calculations throughout this paper. 

Lets start with following general theorem,
\begin{theorem}
Suppose $F :\mathcal{A}\longrightarrow \mathcal{B}$ and $G: \mathcal{B}\longrightarrow \mathcal{C}$ be functors between abelian categories such that $G (K_F) \subset K_G$ (look at the definition of derived functors). Then one gets the following identity,
\begin{equation}\label{functorcombination}
R (G \circ F) = R G \circ R F.
\end{equation}
\end{theorem}
This theorem looks pretty simple, but it allows us to combine derived functors. Basically it says there is a spectral sequence,
\begin{equation}\label{Grothendieckspectralseq}
E_2^{p,q} := R^{p} G(R^q (F)) \Longrightarrow E_{\infty}^{p+q} :=R^{p+q} G \circ F.
\end{equation} 

Here we review some of the applications. First lets consider the direct image of a bounded complex,
\begin{equation}
R^i f_* (\mathcal{H}^j(\mathcal{F}^{\bullet})) \Rightarrow R^{i+j} f_* \mathcal{F}^{\bullet}.
\end{equation}
Obviously one can write a similar spectral sequence formula to compute the derived functor of complexes. Another example is the global section functor over a variety $X$, $\Gamma : Coh(X) \longrightarrow \mathcal{A}b$. The direct images of this functor are just the cohomology of sheaves  \cite{Hartshorne}, i.e. $R^i \Gamma (\mathcal{F}) = H^i(X,\mathcal{F})$. Now let combine this with the direct image functor induced by a proper morphism $f: X \longrightarrow Y$,
\begin{eqnarray}\label{Leray}
&\Gamma_Y : Coh(Y) : \longrightarrow point, \quad \Gamma_X =\Gamma_Y \circ f_* : Coh(X) \longrightarrow point, \nonumber\\
& R\Gamma_X (\mathcal{F}) =R\Gamma_Y \circ Rf_* (\mathbb{F}) ,\nonumber\\
& E_2^{p,q} = H^p(Y, R^qf_*\mathcal{F}) \Rightarrow E_{\infty}^{p+q}= H^{p+q} (X,\mathcal{F}).
\end{eqnarray}
Last line is nothing but Leray spectral sequence. As the final example consider the relation between local extension $\mathcal{E}xt$, and the global extension $Ext$,

 \begin{equation}
 R\Gamma \circ R \mathcal{H}om_{\mathcal{O}_X} (\mathcal{F}^{\bullet},\mathcal{G}^{\bullet}) = R Hom_{D^b(X)} (\mathcal{F}^{\bullet},\mathcal{G}^{\bullet}).
 \end{equation}
 In particular if we apply this to concentrated complexes at zero position (i.e. a single coherent sheaf), we get the following famous result,
 \begin{equation}
 H^i (X,\mathcal{E}xt^j_{\mathcal{O}_X}(\mathcal{F},\mathcal{G})) \Rightarrow Ext^{i+j}_X (\mathcal{F},\mathcal{G})
 \end{equation}

\begin{theorem}[Base change formula]\label{Basechange}  Consider the following commutative diagram of proper morphisms,
\begin{equation}
\begin{tikzcd}
X \arrow[d,"f" '] \arrow[r,"g"] & X' \arrow[d,"f'"] \\
Y \arrow[r, "g'"] & Y'
\end{tikzcd} \nonumber
\end{equation}
Then, in general, there is a morphism of functors ,
\begin{equation}
L f'^* Rg'_* \longrightarrow Rf_* Lg^*.
\end{equation}  
In particular if $f$ ($g$) is flat, then $f'$ ($g'$) is flat, and the above morphism is actually isomorphism of functor. 
\end{theorem}
One of the main properties of Fourier Mukai functor is its compatibility with the base change, and therefore the theorem above will be very useful.

\begin{definition}[Dualizing Complex]\label{DualizingComplex} Consider a proper morphism $f_ X\longrightarrow Y$, it's dualizing complex is defined as,
\begin{equation}
Hom_{D^b(Y)}(Rf_* \mathcal{F}^{\bullet},\mathcal{G}^{\bullet}) = Hom_{D^b(X)} (\mathcal{F}^{\bullet},f^! \mathcal{G}^{\bullet}).
\end{equation}
In particular it satisfies the identities,
\begin{eqnarray}
& f^! \mathcal{G}^{\bullet} = Lf^* \mathcal{G} \otimes^L f^! \mathcal{O}_Y , \\
&\begin{tikzcd}
X \arrow[dr,"h" '] \arrow[r,"f"] & Y \arrow[d,"g"] \\
& Z 
\end{tikzcd} \quad s.t. \quad h= g\circ f  \Longrightarrow h^! = f^! \circ g^! .
\end{eqnarray} 
\end{definition}
So by the first identity one only needs to know the dualizing complex of morphism relative to the structure sheaf. 
\begin{definition}\label{Gorenstein} A morphism is called \emph{Gorenstein} if the dualizing complex is a concentrated complex, i.e. $f^! \mathcal{O}_Y = \Omega {[k]}$ for some $k \in \mathbb{Z}$.
\end{definition} 
There two specific cases that will be useful for us in this paper, 
\begin{description}
\item[Flat Fibration] In this case $f^! \mathcal{O}_Y = \omega_{X/Y}[n] $, where $n$ is the relative dimension (i.e. the dimension of the fibers), and $\omega_{X/Y} = \omega_X \otimes f^* \omega_Y$.
\item[Complete intersection] This is an inclusion morphism $f : X \hookrightarrow Y$ where $X$ is a complete intersection of varieties in $Y$. In this case $f^! \mathcal{O}_Y = det(\mathcal{N}){[-d]}$, where $\mathcal{N}$ is the normal bundle, and $d$ is the codimension of $X$ is $Y$. 
\end{description} 
The definition above is called Grothendieck-Verdier duality, and it is a general form of Serre duality. There is also a local version of this duality,

\begin{equation}
R\mathcal{H}om_{\mathcal{O}_Y} (Rf_* \mathcal{F}^{\bullet},\mathcal{G}^{\bullet}) = Rf_* R\mathcal{H}om_{\mathcal{O}_X} (\mathcal{F}^{\bullet},Lf^* \mathcal{G}^{\bullet}\otimes^L f^! \mathcal{O}_Y).
\end{equation}
\begin{definition} One can define derived dual of a complex $\mathcal{F}^{\bullet}\in D^b(X)$ as,
\begin{equation}
\mathcal{F}^{\bullet \vee} := R\mathcal{H}om_{\mathcal{O}_X} (\mathcal{F}^{\bullet},\mathcal{O}_X).
\end{equation}
\end{definition}
\begin{theorem}
\begin{equation}
R\mathcal{H}om (\mathcal{F}^{\bullet},\mathcal{G}^{\bullet}) \simeq R\mathcal{H}om (\mathcal{O}_X, \mathcal{F}^{\bullet \vee}\otimes^L\mathcal{G}^{\bullet}) \simeq \mathcal{F}^{\bullet \vee}\otimes^L\mathcal{G}^{\bullet}
\end{equation}
\end{theorem}
\begin{theorem}\label{adjunction} $Rf_* \dashv Lf^*$, 
\begin{eqnarray}
& R Hom_{D^b(Y)} (\mathcal{F}^{\bullet},Rf_* \mathcal{G}^{\bullet}) \simeq R Hom_{D^b(X)} (Lf^* \mathcal{F}^{\bullet}, \mathcal{G}^{\bullet}) , \\
& R \mathcal{H}om_{\mathcal{O}_Y} (\mathcal{F}^{\bullet},Rf_* \mathcal{G}^{\bullet}) \simeq Rf_* R \mathcal{H}om_{\mathcal{O}_X} (Lf^* \mathcal{F}^{\bullet}, \mathcal{G}^{\bullet}).
\end{eqnarray}
\end{theorem}
\begin{theorem}[Projection Formula]\label{projection folrmula}
\begin{equation}
R f_* (Lf^* \mathcal{F}^{\bullet} \otimes^L \mathcal{G}^{\bullet}) = \mathcal{F}^{\bullet} \otimes^L R f_*  \mathcal{G}^{\bullet}.
\end{equation}
\end{theorem}

\begin{theorem} From \ref{Basechange}, and the commutative diagram bellow for a \emph{projective morphism} $f$,
\begin{equation}
\begin{tikzcd}
f^{-1}(p) \arrow[d,"f_p" ']\arrow[r, hook, "i_f" ']  & X \arrow[d,"f"] \\
p \arrow[r, hook ,"i" '] & Y
\end{tikzcd} 
\end{equation}
we get the following results when $\mathcal{F}\in Coh(X)$. They will be very useful in many cases, and also give a rather clear intuitive picture about the direct images,
\begin{eqnarray}\label{localformulae}
& Li^* Rf_* \mathcal{F} \longrightarrow R f_{p*} (Li_f^* \mathcal{F}), \nonumber\\
& \phi^j :( L i^* Rf_*  \mathcal{F} )^j = Tor_{-j}^{i^{-1}\mathcal{O}_Y} (Rf_*\mathcal{F} , \mathcal{O}_p) =R^j f_*\mathcal{F} \otimes \mathcal{O}_p  \longrightarrow H^j (f_p^{-1}(p),i_f^* \mathcal{F} ).
\end{eqnarray} 
It is proved in \cite{Hartshorne} III.12.10 that $\phi^j$ is isomorphism if and only of it is surjective, and $R^j f_* \mathcal{F}$ is locally free if and only if $\phi^{j-1}$ is surjective.  
\end{theorem}
\end{labeling}

\section{Integral functors}\label{AppendixB}

In this section we briefly review the main features of integral functors, specially the Fourier Mukai functors which are the important special cases. (For more details, the interested reader can look at \cite{BBRH} and \cite{Huybrechts})
\begin{definition}
Let $D^b(X)$ and $D^b(Y)$ be the derived category of varieties $X$ and $Y$. Consider the following morphisms,

\begin{eqnarray}
&\begin{tikzcd}
 & X \times Y \arrow[dl,"\pi_X "']\arrow[dr,"\pi_Y "] & \\
 X & & Y 
\end{tikzcd}
\end{eqnarray}
Then the integral functor $\Phi_{X \rightarrow Y}^{\mathcal{P}^{\bullet}}$ is defined in the following way, 

\begin{eqnarray}
&\Phi _{X \rightarrow Y}^{\mathcal{P}^{\bullet}} : D^b(X) \longrightarrow D^b(Y), \nonumber \\[10pt]
&\Phi_{X \rightarrow Y}^{\mathcal{P}^{\bullet}} (\dots) := R\pi_{Y*} (\pi_X^* (\dots) \otimes^L \mathcal{P}^{\bullet}),
\end{eqnarray}
where $\pi_X$ and $\pi_Y$ are projections to the corresponding factors, and $\mathcal{P}^{\bullet}$ is the kernel of the transform. Note that $\pi_X$ is a flat morphism, so $L\pi_X^* = \pi_X^*$. \footnote{Such functors are quite similar to the familiar integral transform of functions. Remember that to find the integral transform of  $f(x)$ with $x\in \mathbb{R}^1$ we first consider it as a function in a product space space $\mathbb{R}^1\times \mathbb{R}^1$. This is similar to the pull back $\pi_X^*$ above. Then we multiply $f(x)$ with a kernel $K(x,y)$ which is the function in $\mathbb{R}^1\times \mathbb{R}^1$, this part is similar to the tensor product in the formula above, finally we integrate over $x$, $g(y) = \int dx f(x) K(x,y)$, which is analogues to the push forward $R\pi_{Y*}$. } In particular if the integral transform of a sheaf $\mathcal{E}$ (consider it as complex which is only non-zero at the zero entry, i.e. concentrated on the zero position) is concentrated the $i$th position, it is called a $WIT_i$ sheaf.
\end{definition}

Note that any integral functor is a composition of three exact functors in derived categories, derived inverse image, derived tensor product and derived direct image. So  $\Phi_{X \rightarrow Y}^{\mathcal{P}^{\bullet}} $ is also an exact functor. In particular, to any short exact sequence there is an associated long exact sequence induced by that integral functor.

We are particularly interested in ``relative" integral transforms. Suppose $\Phi_{X\rightarrow Y}^{\mathcal{K}} :D^b(X) \longrightarrow D^b(Y)$ be an integral transform, for any variety $T$, the corresponding relative integral functor (relative to $T$) $\Phi^{\mathcal{K}_T^{\bullet}}_{X\times T \rightarrow Y\times T}$ is defined as

\begin{eqnarray}
&\begin{tikzcd}
 & X \times Y \times T \arrow[dl,"\pi_{X\times T} "']\arrow[dr,"\pi_{Y\times T} "] \arrow[d,"\pi_{X\times Y}"] & \\
 X\times T & X\times Y & Y\times T 
\end{tikzcd} \nonumber \\[10pt]
&\Phi_{T}^{\mathcal{K}_T^{\bullet}} (\dots) := R \pi_{Y\times T _*} (\pi_{X\times T}^{*}(\dots) \otimes^L \mathcal{K}_T^{\bullet}), \nonumber\\[10pt]
&\mathcal{K}_T^{\bullet}:=\pi_{X\times Y}^* \mathcal{K}^{\bullet}. 
\end{eqnarray} 

Now consider a morphism of varieties $f: S\longrightarrow T$, and the induced relative morphisms: $f_X : S \times X \longrightarrow T \times X$ and $f_Y : S \times Y \longrightarrow T \times Y$, then one can prove the following functorial isomorphism,

\begin{equation}
Lf_Y^* \Phi_T (\mathcal{E}^{\bullet}) \simeq \Phi_S (Lf_X^* \mathcal{E}^{\bullet}),
\end{equation}
with $\mathcal{E}^{\bullet} \in D^b(X\times T)$.  In particular if $j_{t}: {t} \longrightarrow T$ be the inclusion of a point $t$, then the identity above gives,

\begin{equation}
Lj_t^{*} \Phi_T(\mathcal{E}^{\bullet}) = \Phi_t(L j_t^* \mathcal{E}^{\bullet}),
\end{equation} 
This has important consequences: first of all if $\mathcal{E}$ is a sheaf, one can prove (by checking the spectral sequences of the combined functors),

\begin{eqnarray}
\Phi^{n_{m}}_{t} (j_t^* \mathcal{E}) \simeq j_t^* \Phi^{n_{m}}_T (\mathcal{E}),
\end{eqnarray}
where $n_m$ is the maximal integer that either $\Phi^{n_{m}}_{t}$ or $\Phi^{n_{m}}_T$ is non zero. Moreover, if both $\mathcal{E}$ and $ \Phi^{i}_T (\mathcal{E})$ are flat over $T$, then $\mathcal{E}_t$ is $WIT_i$ relative to $\Phi_t$ if and only if $\mathcal{E}$ is $WIT_i$ relative to $\Phi_T$.  This is an important point, and when we want to describe the Fourier-Mukai transform of vector bundles which are unstable over some non generic elliptic fibers, or when we need to deal with general coherent sheaves, it is going to help us.

Finally we mention that there are similar result for non trivial fibration, which we discuss briefly later. For now, let's move on to review Fourier-Mukai functors briefly.

\begin{definition}
A Fourier Mukai functor is an integral functor which is also an exact equivalence. 
\end{definition}

Probably the first important point about Fourier-Mukai functors is that any equivalence can be written as Fourier-Mukai,

\begin{theorem}[Orlov's representability theorem] \label{orlov}
Let $X$ and $Y$ be two smooth projective varieties, and let 
\begin{equation}
F : D^b(X) \longrightarrow D^b(Y) \nonumber
\end{equation}
be a fully faithful exact functor. If $F$ admits right and left adjoint functors, then there exists an object $\mathcal{P}^{\bullet} \in D^b(X\times Y)$ unique up to isomorphism such that $F$ is isomorphic to a Fourier Mukai functor $\Phi^{\mathcal{P}}_{X\rightarrow Y}$.
\end{theorem}

There is a partial inverse to this theorem, due to Bondal and Orlov, which states when an integral functor is indeed fully faithful, i.e. it puts constraints over the kernel of the transform,

\begin{theorem}
Let $X$ and $Y$ be smooth projective varieties. Consider $\Phi^{\mathcal{P}^{\bullet}}_{X\rightarrow Y}: D^b(X)\longrightarrow D^b(Y)$ with $\mathcal{P}^{\bullet}$ in $D^b(X\times Y)$. Then $\Phi^{\mathcal{P}^{\bullet}}_{X\rightarrow Y}$ is a fully faithful functor if and only if $\mathcal{P}^{\bullet}$ is a strongly simple object over $X$, i.e. 
\begin{eqnarray}
&&Hom^i_{D^b(Y)}(Lj_{x_1}^* \mathcal{P}^{\bullet},Lj_{x_2}^* \mathcal{P}^{\bullet}) =0 \quad unless \quad x_1=x_2 \quad and \quad 0 \le i \le dimX; \\
&&Hom^0_{D^b(Y)}(Lj_{x}^* \mathcal{P}^{\bullet},Lj_{x}^* \mathcal{P}^{\bullet}) = \mathbb{C}.
\end{eqnarray}
In addition, if $L j_x^* \mathcal{P}^{\bullet}$ is a special object of $D^b(Y)$, i.e. $L j_x^* \mathcal{P}^{\bullet} \otimes K_Y \simeq L j_x^* \mathcal{P}^{\bullet}$, then  $\Phi^{\mathcal{P}^{\bullet}}_{X\rightarrow Y}$ is an equivalence.
\end{theorem}
In particular if both $X$ and $Y$ are both smooth Calabi-Yau varieties, and the kernel is a strongly simple object, then the corresponding integral functor is a Fourier-Mukai functor.

It is worth to mention another very important property of Fourier-Mukai functors, and that is these kind of integral functors are sensitive to smoothness and `` Calabi-Yau ness", and dimension. In other words, if tow varieties $X$ and $Y$ are Fourier-Mukai partners (their derived category are equivalent), then $X$ is smooth if and only if $Y$ is smooth (this proved by Serre's criterion on regular local rings of finite homological dimension), and $X$ is Calabi-Yau if and only if $Y$ is Calabi-Yau (this is proved by using Grothendieck-Verdier duality), and both of them must have the same dimension. There are also other geometrical constraints which are induced by the equivalence condition, but we ignore them here. 

We finish this section by quickly deriving the inverse transform of a Fourier-Mukai functor $\Phi_{X\rightarrow Y}^{\mathcal{P}^{\bullet}}$. Since for an equivalence of categories, the adjoint functor is actually the inverse functor, one can find it easily for the Fourier Mukai functor as follows,

\begin{eqnarray}
RHom_{D^b(Y)} (\Phi_{X\rightarrow Y}^{\mathcal{P}^{\bullet}}(\mathcal{F}^{\bullet}), \mathcal{G}^{\bullet}) &=&  RHom_{D^b(X\times Y)}(\pi_X^*\mathcal{F}^{\bullet},\pi_Y^*\mathcal{G}^{\bullet} \otimes^L \mathcal{P}^{\bullet \vee} \otimes\pi_X^*\omega_X[n]) \nonumber \\
&=& RHom_{D^b(X)} (\mathcal{F}^{\bullet}, R\pi_{X*} (\pi_Y^* \mathcal{F}^{\bullet} \otimes^L \mathcal{P}^{\bullet \vee}\otimes \pi_X^*\omega_X[n] )) \\
&=& RHom_{D^b(X)}(\mathcal{F}^{\bullet},\Phi_{Y\rightarrow X}^{\mathcal{P}^{\bullet \vee}\otimes \pi_X^* \omega_X [n]} (\mathcal{G}^{\bullet})),\nonumber
\end{eqnarray}
where $\mathcal{F}^{\bullet}$ and $\mathcal{G}^{\bullet}$ are generic objects of derived category of varieties $X$ and $Y$, $n$ is the dimension of both $X$ and $Y$\footnote{Actually uniqueness of the inverse implies the dimension of $X$ and $Y$ must be the same.}, and $\omega_X$ is the canonical sheaf of $X$. Therefor the ``inverse transform" is itself a Fourier Mukai functor,
\begin{eqnarray}
\Phi_{Y\rightarrow X}^{\mathcal{P}^{\bullet \vee}\otimes \pi_X^* \omega_X [n]}.
\end{eqnarray}

\newpage

\bibliographystyle{utphys}
\bibliography{refs}

\end{document}